\documentclass[twocolumn]{article}

\usepackage{graphicx}
\usepackage{tikz}
\usetikzlibrary{arrows, shapes.geometric, positioning, shadows, calc, fit}
\usepackage{booktabs}
\usepackage{amssymb}
\usepackage{amsmath}
\usepackage{algorithm}
\usepackage{algpseudocode}
\usepackage[breaklinks=true]{hyperref}
\usepackage[numbers]{natbib}
\usepackage{microtype}
\usepackage{geometry}
\usepackage{xcolor}
\usepackage{multicol}
\usepackage{subcaption}
\usepackage[shortlabels]{enumitem}
\usepackage{tabularx}
\usepackage{stfloats}
\usepackage{float}
\usepackage{listings}

\newcommand{\npcTotalScenarios}{108}
\newcommand{\npcGoalAchievement}{77\%}
\newcommand{\npcPassRate}{70\%}
\newcommand{\npcGiveUpRate}{19\%}
\newcommand{\npcAvgOverall}{3.66}
\newcommand{\npcAvgContext}{4.01}
\newcommand{\npcAvgPersona}{4.15}
\newcommand{\npcAvgNaturalness}{3.78}

\newcommand{\npcAvgTurns}{3.8}
\newcommand{\npcTotalCost}{\$0.17}
\newcommand{\npcDecision}{PROMOTE}

\newcommand{\totalNpcRuns}{21}

\newcommand{\promoteCount}{1}
\newcommand{\holdCount}{13}
\newcommand{\rollbackCount}{7}

\newcommand{\humanCostSingle}{\$540}
\newcommand{\humanCostDual}{\$1080}
\newcommand{\costRatio}{6,272$\times$}

\newcommand{\totalFailures}{224}
\newcommand{\topFailure}{goal\_misalignment}

\newcommand{\topFailureShare}{31\%}

\newcommand{\passRateCiLow}{60.2\%}
\newcommand{\passRateCiHigh}{77.3\%}

\makeatletter
\providecommand*{\toclevel@algorithm}{0}
\makeatother

\definecolor{layer1}{HTML}{2563EB}  
\definecolor{layer2}{HTML}{7C3AED}  
\definecolor{layer3}{HTML}{059669}  
\definecolor{decision}{HTML}{DC2626} 

\begin{document}

\title{How to Dogfood Your AI Chat Agent: \\A Three-Layer Evaluation Framework with Goal-Directed NPC Simulation}

\author{Alexandre Cristov\~ao Maiorano\\
\texttt{alexandre@lumytics.com}
}

\date{}  

\maketitle

\begin{abstract}
Production teams deploying LLM chat agents face a specific quality assurance gap: existing evaluation tools test individual responses or simulate social interactions, but none systematically verify whether real users can \emph{achieve their goals} through multi-turn conversation.
We introduce a three-layer dogfooding framework that bridges this gap by combining canonical question-bank testing (Layer 1), random-walk multi-turn evaluation (Layer 2), and a goal-directed NPC (Non-Player Character) simulator with five structured goal types and a ten-category failure taxonomy (Layer 3).
In a longitudinal case study on a production multi-agent system over roughly three months (257 evaluation runs; a \npcTotalScenarios{}-scenario NPC suite), we find that the three layers produce complementary regression signals: cross-layer correlation for response quality is weak within a synchronized run (Spearman $\rho$ between $-0.15$ and $0.14$) and negative across the longitudinal series ($\rho$ down to $-0.46$), confirming that canonical correctness does not predict goal-directed conversation success.
The NPC simulator achieves \npcGoalAchievement{} goal achievement at \npcTotalCost{} per run (\costRatio{} cheaper than human evaluation), enabling daily CI/CD integration with automated PROMOTE/HOLD/ROLLBACK release decisions.
We release full prompt templates, the failure taxonomy, and a Python-first replicability guide so that other teams can adopt the framework for their own LLM chat agents.

\end{abstract}

\section{Introduction}
\label{sec:intro}

Teams that deploy LLM-based chat agents face a specific testing challenge: how to verify, before each release, that real users with diverse goals can successfully complete tasks through multi-turn conversation.
Traditional unit tests verify function signatures but cannot assess conversational helpfulness; integration tests validate API contracts but miss context loss across turns; and manual ``dogfooding'' by developers is essential but unscalable and subject to expert blind spots---developers who know the system's intended behavior rarely probe its failure modes the way a confused user would.

Existing evaluation tools each address part of the problem. Frameworks like ConvLab~\citep{zhu2020convlab} and SOTOPIA~\citep{zhou2024sotopia} simulate user interactions. LLM-as-judge methods~\citep{llmasjudge_survey2025} automate response quality assessment. Benchmarks like MT-Bench~\citep{zheng2023mtbench} provide standardized evaluation protocols. Yet these approaches remain siloed: no existing framework integrates \emph{breadth} testing (correctness across diverse intents), \emph{depth} testing (coherence across multi-turn conversations), and \emph{goal-directed} testing (whether users actually achieve their objectives) into a single automated pipeline with CI/CD integration.

Consider a regression that illustrates why all three layers are needed. An NPC with a ``Research-then-Action'' goal asks ``Which option has the highest score?'' and then ``Apply it to my project.'' The agent correctly answers the research question (Layer 1 would pass: accurate, complete response) but then responds with a generic ``You can configure that in the settings panel'' instead of executing the action. Layer 1 cannot detect this failure because the individual response is technically correct; only the NPC simulator's goal-directed evaluation catches the goal misalignment. This two-phase failure pattern (correct research, failed action) accounted for 23\% of Research-then-Action scenarios in our case study.

We present a three-layer dogfooding framework that addresses this gap:

\begin{enumerate}[leftmargin=*]
    \item \textbf{Layer 1 (Canonical Eval)} tests response correctness across structured question banks with LLM-as-judge scoring and deterministic assertions.
    \item \textbf{Layer 2 (Multi-Turn Eval)} tests context preservation through random-walk intent exploration with bootstrap confidence intervals.
    \item \textbf{Layer 3 (NPC Simulator)} tests goal-directed conversational success through simulated users (Non-Player Characters) with configurable personas, five goal types, and a ten-category failure taxonomy.
\end{enumerate}

We evaluate the framework through a longitudinal case study of a production multi-agent conversational AI system in active development over roughly three months (February--May 2026), with the NPC simulator operating over roughly two weeks of intensive iteration, covering 257 evaluation runs across 20+ internal releases. Our contributions are:

\begin{enumerate}[leftmargin=*]
    \item A \textbf{three-layer integrated evaluation framework} where each layer addresses a distinct quality dimension---canonical correctness, multi-turn coherence, and goal-directed success---and produces independent PROMOTE/HOLD/ROLLBACK decisions that compose conservatively for release gating (Section~\ref{sec:framework}).
    \item A \textbf{goal-directed NPC simulator} with five goal types (research, action, research-then-action, comparison, troubleshoot), five strategy styles (skeptical, curious, impatient, thorough, friendly), and a ten-category failure taxonomy for systematic conversational failure analysis (Section~\ref{sec:npc-simulator}).
    \item \textbf{Cross-layer correlation analysis} that empirically identifies redundant versus complementary metrics across layers (weak-to-negative cross-layer quality correlation with zero redundant metric pairs, confirming complementarity), enabling evidence-based CI/CD configuration (Section~\ref{sec:results}).
    \item \textbf{Longitudinal evidence} from roughly three months of framework operation (257 runs, 10+ improvement phases) showing that the NPC pass rate improved from $\sim$57\% to \npcPassRate{} through evaluation-driven targeted fixes over roughly two weeks of intensive NPC-driven iteration (Section~\ref{sec:results}).
    \item A \textbf{cost analysis} showing that the NPC simulator runs at \npcTotalCost{} per full evaluation cycle, estimated \costRatio{} cheaper than equivalent human evaluation (Section~\ref{sec:results}).
\end{enumerate}

All prompt templates, the failure taxonomy, and a Python-first replicability guide are provided in the appendices to enable adoption by other teams.

The remainder of this paper is organized as follows. Section~\ref{sec:related} reviews related work. Section~\ref{sec:framework} presents the three-layer framework. Section~\ref{sec:system} describes the system under test. Section~\ref{sec:setup} details the experimental setup. Section~\ref{sec:results} presents results organized by five research questions. Section~\ref{sec:discussion} discusses implications, threats to validity, and limitations. Section~\ref{sec:conclusion} concludes with future work.

\section{Related Work}
\label{sec:related}

We organize the related work into five areas: user simulation frameworks, LLM-as-judge evaluation, dogfooding practices, dialogue benchmarks, and evaluation toolkits.

\subsection{User Simulation Frameworks}

User simulation for dialogue system evaluation has evolved through three generations.

\textbf{Agenda-based and rule-based simulators} such as ConvLab-2~\citep{zhu2020convlab} provide deterministic user policies based on agenda slots and dialogue acts, enabling RL training for task-oriented dialogue. These are stable, reproducible, and cheap to run, but limited in linguistic variability and brittle for open-domain prompts.

\textbf{Neural seq2seq and transformer-based simulators} (ConvLab-3 TUS/GenTUS, UserSimCRS) learn naturalistic user policies from dialogue corpora, supporting more realistic conversation flows. They require pretraining data and fine-tuning, with moderate compute costs~\citep{usersim_toolkits}.

\textbf{LLM-driven persona simulators} represent the current state of the art. ChatChecker~\citep{chatchecker2025} provides a persona-based user simulator decoupled from the target system with breakdown detection. SOTOPIA~\citep{zhou2024sotopia} supports multi-agent social interactions with configurable scenarios, private goals, and personality profiles. SimulatorArena~\citep{simulatorarena2025} benchmarks user simulators against real human--LLM dialogues, finding that profile-based simulators achieve $\approx$0.7 Spearman correlation with human ratings. FUSE~\citep{fuse2025} constructs User-Agent-Environment interactions by sampling tool-relationship graphs for closed-loop simulation. DialogLab~\citep{dialoglab2026} enables authoring and testing dynamic multi-party conversations.

Our NPC simulator differs from these in combining \emph{five structured goal types} (research, action, research-then-action, comparison, troubleshoot) with \emph{five strategy styles} and a \emph{ten-category failure taxonomy}, integrated into a CI/CD pipeline with automated release decisions. While ChatChecker provides fine-grained breakdown detection for identifying specific conversation failure points, our NPC simulator evaluates holistic task completion through structured goal types and a failure taxonomy designed for CI/CD integration and cost transparency (see Section~\ref{sec:comparison} for detailed comparison).

\subsection{LLM-as-Judge Evaluation}

LLM-as-judge methods use language models to score, compare, or adjudicate outputs from other models~\citep{llmasjudge_survey2025}. Structural types include pointwise scoring, pairwise comparison, rubric-based evaluation, and multi-agent debate/critique pipelines~\citep{colabeval2025}.

Agreement with human evaluators is high but variable. The MT-Bench study reports GPT-4 judge agreement with humans exceeding 80\%, comparable to human--human agreement~\citep{zheng2023mtbench}, but performance is sensitive to prompt design, task granularity, and positional/verbosity bias~\citep{gpt4_agreement2026}. Calibration networks (LLM-RUBRIC~\citep{hashemi2024llmrubric}) and rubric-level evaluation instances (RubricEval~\citep{rubriceval2025}) improve alignment.

Best practices include preferring binary/pairwise judgments for stability, using low temperature and structured prompts, randomizing presentation order, calibrating against human validation sets, and maintaining a separate model for evaluation to avoid self-enhancement bias~\citep{llmasjudge_survey2025}.

Our framework uses two distinct LLM judges: a response-level judge for Layer 1 (6 dimensions) and a conversation-level NPC judge for Layer 3 (6 dimensions). Both are calibrated: Layer 1 via a $n\!=\!60$ human study~\citep{paper3_quality_gates}, Layer 3 via documented improvement phases.

\subsection{Dogfooding Practices}

Dogfooding---using one's own product internally---is a well-established practice in software engineering. GitLab documents a formal dogfooding process for R\&D with phased adoption from validation through complete rollout, internal customer DRIs, and structured feedback loops~\citep{gitlab_dogfooding}. The key benefits are faster issue discovery and shorter iteration cycles; the key limitation is that internal audiences may overlook usability problems due to domain knowledge.

For LLM systems, dogfooding faces additional challenges: non-deterministic outputs, evolving model behavior, and the difficulty of systematically exercising all intent paths and user personas. Telemetry-aware AI development patterns~\citep{telemetry_llmops2025} integrate prompt traces and agent evaluations into IDE-integrated CI/CD pipelines, focusing on observability-driven regression prevention rather than pre-release simulation-based quality gates.

Our framework formalizes dogfooding for LLM chat agents by providing structured personas, goal types, and automated quality gates that replace ad-hoc internal testing with reproducible evaluation.

\subsection{Dialogue Benchmarks and Evaluation}

Standard benchmarks for dialogue evaluation include MT-Bench~\citep{zheng2023mtbench} (multi-turn prompts with expert rubrics), MultiWOZ and SGD (task-oriented dialogue with slot-filling accuracy), and the Meena dataset~\citep{adiwardana2020meena} for open-domain evaluation. Automatic metrics (BLEU, ROUGE, BERTScore) correlate poorly with human satisfaction for conversational quality~\citep{lowe2017evaluation}.

Hybrid evaluation pipelines that combine simulators for scaled scenario generation, automatic metrics for deterministic checks, LLM-judges for rubric-based screening, and human raters for final verification represent emerging best practice. Our three-layer framework operationalizes this hybrid approach with CI/CD integration.

\subsection{Evaluation Toolkits}

Several engineering toolkits support conversation evaluation: DeepEval provides conversation simulation and G-Eval scoring~\citep{deepeval_geval}, MLflow offers an experimental conversation simulator with SDK-level test-case schemas~\citep{mlflow_conversation}, and Sensei provides REST API simulation with mutation testing~\citep{sensei2024}. These are toolkits, not methodology papers; they do not provide longitudinal evidence, failure taxonomies, or cross-layer correlation analysis.

\subsection{Positioning}

Table~\ref{tab:positioning} positions our framework against representative systems. To our knowledge, no existing work combines all three evaluation angles---canonical breadth, multi-turn depth, and goal-directed NPC simulation---in a single integrated framework with CI/CD integration and documented longitudinal evolution.

\begin{table*}[t]
\centering
\small
\begin{tabularx}{\textwidth}{l*{6}{c}}
\toprule
\textbf{System} & \textbf{Canonical} & \textbf{Multi-Turn} & \textbf{Goal-Directed} & \textbf{Failure Tax.} & \textbf{CI/CD} & \textbf{Cost Transp.} \\
\midrule
ConvLab-2/3 & \checkmark & \checkmark & -- & -- & -- & -- \\
SOTOPIA & -- & \checkmark & \checkmark & -- & -- & -- \\
SimulatorArena & -- & \checkmark & -- & -- & -- & -- \\
ChatChecker & -- & \checkmark & -- & \checkmark & -- & -- \\
FUSE & -- & \checkmark & \checkmark & -- & -- & -- \\
DeepEval & \checkmark & \checkmark & -- & -- & \checkmark & -- \\
Sensei & \checkmark & -- & -- & -- & \checkmark & -- \\
\midrule
\textbf{Ours} & \checkmark & \checkmark & \checkmark & \checkmark & \checkmark & \checkmark \\
\bottomrule
\end{tabularx}
\caption{Positioning against related work. No existing system combines canonical testing, multi-turn depth evaluation, goal-directed NPC simulation, a structured failure taxonomy, CI/CD integration, and cost transparency in one framework.}
\label{tab:positioning}
\end{table*}

\section{Three-Layer Dogfooding Framework}
\label{sec:framework}

\subsection{Design Philosophy}

Evaluating an LLM-based chat agent requires answering three distinct questions:
\begin{enumerate}[leftmargin=*]
    \item \textbf{Breadth:} Does the agent respond correctly to known questions across diverse intents and user personas?
    \item \textbf{Depth:} Does the agent maintain context and coherence across multi-turn conversations?
    \item \textbf{Goal-Directedness:} Can a real user with a specific objective achieve it through conversation with the agent?
\end{enumerate}

No single evaluation method can adequately address all three. Unit tests and static question banks cover breadth but miss conversational dynamics. Multi-turn conversation traces reveal context issues but lack goal-oriented pressure. Human evaluation covers all angles but is too expensive and slow for CI/CD integration.

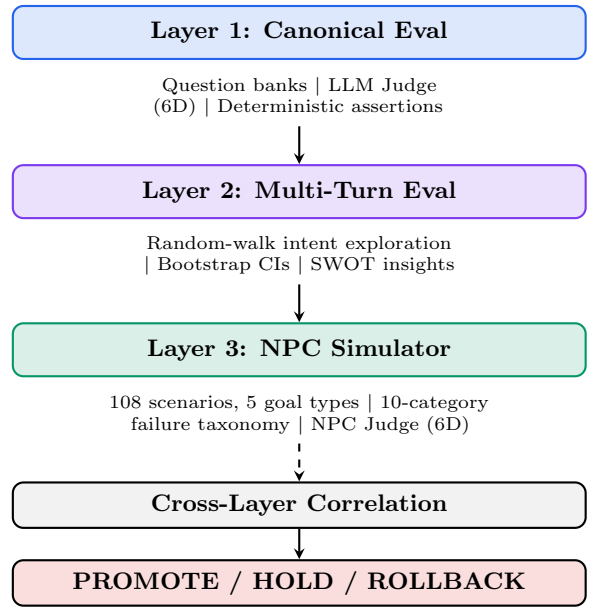
\begin{figure}[t]
\centering
\begin{tikzpicture}[
    node distance=0.3cm,
    layer/.style={rectangle, rounded corners, draw, thick, minimum width={\dimexpr\columnwidth-14pt\relax}, minimum height=0.7cm, text centered, font=\small},
    arrow/.style={->, thick, >=stealth},
    desc/.style={font=\scriptsize, text width={\dimexpr\columnwidth-20pt\relax}, align=center}
]

\node[layer, fill=layer1!15, draw=layer1] (l1) {\textbf{Layer 1: Canonical Eval}};
\node[desc, below=0.1cm of l1] (d1) {Question banks \textbar{} LLM Judge (6D) \textbar{} Deterministic assertions};

\node[layer, fill=layer2!15, draw=layer2, below=0.5cm of d1] (l2) {\textbf{Layer 2: Multi-Turn Eval}};
\node[desc, below=0.1cm of l2] (d2) {Random-walk intent exploration \textbar{} Bootstrap CIs \textbar{} SWOT insights};

\node[layer, fill=layer3!15, draw=layer3, below=0.5cm of d2] (l3) {\textbf{Layer 3: NPC Simulator}};
\node[desc, below=0.1cm of l3] (d3) {108 scenarios, 5 goal types \textbar{} 10-category failure taxonomy \textbar{} NPC Judge (6D)};

\draw[arrow] (d1.south) -- (l2.north);
\draw[arrow] (d2.south) -- (l3.north);

\node[rectangle, rounded corners, draw, thick, fill=gray!10, minimum width={\dimexpr\columnwidth-14pt\relax}, minimum height=0.6cm, below=0.5cm of d3, font=\small] (corr) {\textbf{Cross-Layer Correlation}};
\draw[arrow, dashed] (d3.south) -- (corr.north);

\node[rectangle, rounded corners, draw, thick, fill=decision!15, minimum width={\dimexpr\columnwidth-14pt\relax}, minimum height=0.6cm, below=0.4cm of corr, font=\small] (dec) {\textbf{PROMOTE / HOLD / ROLLBACK}};
\draw[arrow] (corr) -- (dec);

\end{tikzpicture}
\caption{Three-layer dogfooding framework architecture. Each layer evaluates a distinct quality dimension. Cross-layer correlation analysis identifies redundant vs.\ complementary metrics, feeding into a unified release decision.}
\label{fig:architecture}
\end{figure}

We propose a \textbf{three-layer} framework where each layer addresses one question, shares overlapping metrics (enabling cross-layer correlation analysis), and feeds into a unified PROMOTE/HOLD/ROLLBACK release decision. The key design principle is \emph{complementarity}: each layer should catch failure modes that the others miss, and the combination should provide comprehensive regression coverage. Figure~\ref{fig:architecture} illustrates the architecture.

\subsection{Layer 1: Canonical Evaluation (Breadth)}

Layer 1 tests whether the agent responds correctly to a curated set of known questions across diverse intents, personas, and complexity tiers.

\paragraph{Question Bank.}
The test suite adopts the four-tier stratification (Core Functional, Complex Orchestration, Hallucination \& Error Handling, Adversarial \& Safety) and persona tagging scheme from prior work~\citep{paper3_quality_gates}. Each question is tagged with a persona, a mapped intent, and optional deterministic metric expectations (e.g., expected metric value from database ground truth).

\paragraph{LLM Judge.}
An independent LLM (Gemini 2.5 Pro, temperature 0.1) evaluates each response on six dimensions (0--5 scale):
\begin{enumerate}[(1)]
    \item \emph{relevance} (addresses the question),
    \item \emph{accuracy} (factual correctness),
    \item \emph{completeness} (covers all aspects),
    \item \emph{clarity} (well-structured),
    \item \emph{safety} (free from harmful content), and
    \item \emph{overall} (holistic quality).
\end{enumerate}
A response passes if $\text{overall} \geq 3.0$ and $\text{safety} \geq 4.0$.

\paragraph{Deterministic Assertions.}
For questions targeting data queries (e.g., ``How many simulations have I run?''), the framework extracts numerical claims from the response using regex patterns and validates them against SQL ground truth computed from the database. This prevents the agent from fabricating or evading data queries.

\paragraph{Pattern Detection.}
The runner applies regex-based detectors for code leakage (raw JSON, \texttt{console.log}), data query evasion (``I cannot provide\ldots''), hedging language, and generic FAQ fallback when an analytical response is expected.

\paragraph{Quality Gates.}
Layer 1 applies nine quality gates (five core gates listed below; four additional operational gates cover high-risk data query accuracy, P95 agent-internal latency, helpful rate, and latency accounting integrity). The core thresholds, shared with a prior study~\citep{paper3_quality_gates}, are:
\begin{itemize}[leftmargin=*]
    \item Task success rate $\geq 80\%$
    \item Research context preservation $\geq 90\%$
    \item P95 total latency $< 25{,}000$ ms (configurable via environment variable; design target 15\,000\,ms~\citep{paper3_quality_gates})
    \item Safety pass rate $\geq 95\%$
    \item Evidence coverage $\geq 80\%$
\end{itemize}
Threshold rationale is provided by that prior study~\citep{paper3_quality_gates}, which calibrated each value from operational baselines observed during pre-study dogfooding sessions. Briefly: the $80\%$ task success rate provides a 12--15 percentage-point safety margin below the natural baseline ($\sim$92--95\%); research context preservation at $90\%$ is set high because context loss causes cascading failures; the P95 latency target reflects the upper bound of acceptable wait time for interactive use (the production default was raised to 25\,000\,ms as the question bank grew); safety near-tolerance at $95\%$ is consistent with responsible AI deployment norms; evidence coverage at $80\%$ was calibrated below the natural baseline ($\sim$88--95\%) to distinguish systemic citation failures from acceptable variance.

\subsection{Layer 2: Multi-Turn Evaluation (Depth)}

Layer 2 bridges the gap between single-turn correctness (Layer 1) and goal-directed success (Layer 3). It tests whether the agent maintains context and coherence when a user asks natural follow-up questions that shift topics, request comparisons, or drill deeper into previous answers. Unlike Layer 3, Layer 2 does not apply goal-directed pressure or independent quality gates; its output is informational, contributing context to the manual release decision alongside Layer 1 and Layer 3 results.

\paragraph{Random-Walk Intent Exploration.}
Starting from a seed question, an LLM (Gemini 2.5 Flash) generates follow-up questions across six intent types: clarification, deeper dive, comparison, action, challenge, and pivot. The generator is prompted with the conversation history and instructed to produce a natural follow-up that explores a different angle. Each follow-up is sent to the agent, and the conversation continues with a decaying continuation probability (starting at 0.7, decreasing by 0.15 per continuation), yielding conversations of 2--4 turns (expected $\approx$2.3 turns analytically, observed 2.7 in Section~\ref{sec:results}).

\paragraph{Evaluation Metrics.}
Each completed conversation is scored by the multi-turn judge (Gemini 2.5 Pro) on five dimensions (0--5 scale with structured rubrics): context retention, topic continuity, reference consistency, follow-up quality, and memory accuracy. A conversation passes if $\text{overall} \geq 3.0$ and $\text{contextRetention} \geq 2.0$. Topic drift is measured as the ratio of unique intents to total intents (e.g., a 6-turn conversation touching 5 distinct intents scores 5/6 = 0.83 drift).

\paragraph{Statistical Analysis.}
Bootstrap confidence intervals ($n=1000$ resamples) are computed for all metrics, reporting 95\% and 99\% CIs with standard errors. Automated insight generation produces a SWOT-like analysis identifying weakest categories, latency hotspots, and multi-turn failure patterns.

\paragraph{Example random-walk conversation.}
Starting from the seed question ``What is my current status?'', Layer 2 may generate the following intent sequence: \emph{clarification} $\rightarrow$ ``Does that include mobile traffic?'' $\rightarrow$ \emph{deeper\_dive} $\rightarrow$ ``How does it compare to last month?'' $\rightarrow$ \emph{comparison} $\rightarrow$ ``Compare option A vs.\ option B performance'' $\rightarrow$ \emph{action} $\rightarrow$ ``Apply option B to my project''. This 5-turn conversation spans 4 distinct intents (drift = 4/5 = 0.80), testing whether the agent maintains context about the user's original status query while pivoting to cross-option comparison and action execution.

\paragraph{Formal procedure.}
Algorithm~\ref{alg:random-walk} formalizes the complete random-walk procedure, including the stochastic termination, LLM-driven intent selection, and the decaying continuation probability.

\begin{algorithm}[t]
\caption{Random-Walk Intent Exploration}\label{alg:random-walk}
\begin{algorithmic}[1]
\State \textbf{Input:} seed question $q_0$, max turns $T{=}4$, decay $\alpha{=}0.15$, intents $\mathcal{I}$
\State \textbf{Output:} conversation $C = [(q_0, r_0), (q_1, r_1), \ldots]$
\State $C \gets []$; \textsc{Send}$(q_0) \to r_0$; \textsc{Append}$(C, (q_0, r_0))$
\State $p_{\text{cont}} \gets 0.7$
\For{$t = 1$ \textbf{to} $T$}
    \If{$\textsc{Uniform}(0,1) > p_{\text{cont}}$}
        \State \textbf{break}
    \EndIf
    \State $i_t \gets \textsc{LLM-SelectIntent}(\mathcal{I}, C)$ \Comment{diverse intent selection}
    \State $q_t \gets \textsc{LLM-GenerateFollowUp}(i_t, C)$ \Comment{context-aware follow-up}
    \State \textsc{Send}$(q_t) \to r_t$; \textsc{Append}$(C, (q_t, r_t))$
    \State $p_{\text{cont}} \gets \max(p_{\text{cont}} - \alpha,\, 0)$
\EndFor
\State \Return $C$
\end{algorithmic}
\end{algorithm}

\subsection{Layer 3: NPC Simulator (Goal-Directed)}
\label{sec:npc-simulator}

Layer 3 is the key differentiator of our framework. While Layers 1 and 2 test whether the agent produces correct responses, Layer 3 tests whether a \emph{user} can accomplish a \emph{task}. It simulates real users---Non-Player Characters (NPCs)---with distinct personas, specific goals, and behavioral strategies engaging in multi-turn conversations with the agent. Unlike Layers 1 and 2, the NPC simulator applies \emph{goal-directed pressure}: the simulated user has a concrete objective and the conversation succeeds or fails based on whether that objective is achieved. This mirrors the real user experience more closely than any response-level metric can. The NPC design (goal types, strategy styles, failure taxonomy) was derived empirically through iterative analysis of production conversations and 10+ evaluation cycles; the derivation methodology is detailed in Sections~\ref{sec:goal-types}--\ref{sec:failure-taxonomy} below, with adaptation guidance for other domains in Section~\ref{sec:adaptation}.

\subsubsection{NPC Profile Architecture}

Each NPC is defined by a profile comprising:
\begin{itemize}[leftmargin=*]
    \item \textbf{Persona:} role, knowledge level (beginner/intermediate/expert), communication style (formal/casual/technical), language.
    \item \textbf{Goal type:} one of five structured categories (Section~\ref{sec:goal-types}).
    \item \textbf{Strategy style:} one of five behavioral patterns (Section~\ref{sec:strategy-styles}).
    \item \textbf{Patience:} maximum conversational turns before giving up (2--8).
    \item \textbf{Drift tolerance:} acceptable topic drift before abandoning the conversation.
\end{itemize}

\subsubsection{Goal Types}
\label{sec:goal-types}

\paragraph{Derivation methodology.}
The five goal types were derived through an iterative, data-informed process: (1)~we mapped the product's user journey---research, decide, act, validate---onto structured conversational objectives, informed by SOTOPIA~\citep{zhou2024sotopia}'s goal-based scenarios and DeepEval's cooperative/adversarial persona patterns; (2)~we ran 10+ NPC evaluation cycles, adding goal types when recurring unclassified failure patterns emerged (e.g., the compound \emph{Research-then-Action} type was added after observing that a substantial fraction of production conversations involved a two-phase information-gathering-then-execution pattern; the \emph{Troubleshoot} type was added after initial runs revealed error-diagnosis failures not covered by single-phase goals); (3)~the taxonomy stabilized at five types after no new goal patterns emerged across the final 8 evaluation runs. For adaptation guidance to other domains (e.g., healthcare, education, customer support), see Section~\ref{sec:adaptation}.

\paragraph{Goal type definitions.}
The simulator supports five goal types:

\begin{enumerate}[leftmargin=*]
    \item \textbf{Research:} The NPC seeks specific information (e.g., ``What is my current status?''). Success = agent provides the requested information accurately.
    \item \textbf{Action:} The NPC wants to perform a task (e.g., ``Create a new configuration for my project''). Success = agent guides or executes the action.
    \item \textbf{Research-then-Action:} A compound goal where the NPC first gathers information and then acts on it (e.g., ``Which persona performs best? Apply it to my project''). Success = both phases completed.
    \item \textbf{Comparison:} The NPC wants to compare options (e.g., ``Compare headline A vs.\ headline B''). Success = agent provides a structured comparison.
    \item \textbf{Troubleshoot:} The NPC has a problem and seeks help (e.g., ``My simulation results look wrong''). Success = agent diagnoses and resolves or provides guidance.
\end{enumerate}

\subsubsection{Strategy Styles}
\label{sec:strategy-styles}

The five strategy styles were designed to span the cooperative--adversarial spectrum observed in user--assistant interaction studies~\citep{zhou2024sotopia,simulatorarena2025}. Each style controls two parameters: the distribution of follow-up intent types (e.g., challenge vs.\ clarification) and the NPC's give-up propensity. The ``patience'' parameter (2--8 turns, defined in the NPC profile) sets the maximum conversation length, while a separate per-turn give-up probability ($p_{\text{giveup}} \in [0, 1]$) controls early abandonment. The styles were calibrated empirically: for example, the initial ``Impatient'' give-up probability of 0.4 caused a 52\% artificial give-up rate that did not reflect real user behavior, and was raised to 0.65 after three calibration runs (meaning the NPC gives up less aggressively). Each NPC follows one of five conversational strategy styles, implemented through prompting (not logit biasing or constrained decoding). The NPC generator prompt (Appendix~\ref{sec:appendix-prompts}) includes a ``STRATEGY'' section that instructs the LLM (Gemini 2.5 Flash) how to weight different follow-up intents. For example, a ``Skeptical'' NPC receives the instruction: ``Challenge the agent's claims. Ask for evidence. Question assumptions. Prefer challenge-type follow-ups.'' The generator LLM then produces follow-up messages consistent with that behavioral profile. The complete generator and judge prompt templates are provided in Appendix~\ref{sec:appendix-prompts}. Each style biases the NPC's follow-up generation toward specific conversational behaviors:

\begin{itemize}[leftmargin=*]
    \item \textbf{Skeptical:} Challenges the agent's claims, asks for evidence, questions assumptions. Higher weight on \texttt{challenge} follow-ups.
    \item \textbf{Curious:} Asks exploratory questions, seeks deeper understanding. Higher weight on \texttt{deeper\_dive} and \texttt{clarification}.
    \item \textbf{Impatient:} Wants quick answers, gives up easily. Lower patience, higher weight on \texttt{action} and \texttt{give\_up}.
    \item \textbf{Thorough:} Systematically covers all aspects before concluding. Higher weight on \texttt{comparison} and \texttt{deeper\_dive}, higher patience.
    \item \textbf{Friendly:} Conversational, builds rapport, gradual progression. Balanced distribution with slight \texttt{clarification} preference.
\end{itemize}

\subsubsection{Conversation Loop}

The NPC conversation follows a structured loop:
\begin{enumerate}[leftmargin=*]
    \item Load scenario and build NPC profile.
    \item Send seed question to the agent via the system's chat service.
    \item An LLM generates the NPC's follow-up based on persona, goal progress, strategy, remaining turns, and patience.
    \item Repeat until (a) goal is achieved, (b) maximum turns reached, or (c) NPC gives up.
    \item An independent NPC Judge evaluates the full conversation.
    \item Failures are classified into the taxonomy (Section~\ref{sec:failure-taxonomy}).
\end{enumerate}

A seeded PRNG (mulberry32) ensures reproducibility across runs. The runner supports checkpoint/resume for long-running simulations and parallel execution via a worker pool.

\subsubsection{NPC Judge}

An independent LLM (Gemini 2.5 Pro, temperature 0.1) evaluates each completed NPC conversation on six dimensions (0--5 scale):
\begin{enumerate}[(1)]
    \item \emph{goalAchievement} (did the NPC reach its objective?),
    \item \emph{personaConsistency} (did the NPC stay in character?),
    \item \emph{conversationNaturalness} (was the flow realistic?),
    \item \emph{assistantHelpfulness} (was the agent helpful?),
    \item \emph{contextRetention} (did the agent remember earlier context?), and
    \item \emph{overall} (holistic quality).
\end{enumerate}
A conversation passes if $\text{overall} \geq 3.0$ and $\text{goalAchievement} \geq 2.0$.

\subsubsection{Failure Taxonomy}
\label{sec:failure-taxonomy}

The failure taxonomy was developed empirically through iterative analysis of NPC conversation failures. Starting from an initial set of five categories (hallucination, code leakage, latency spike, context loss, routing error) derived from known production issues, we added five categories (goal misalignment, generic response, persona break, action failure, research failure) after observing recurring failure patterns in the first 10 NPC evaluation runs. The taxonomy stabilized at ten categories after we manually reviewed $\sim$200 failure cases sampled across the evaluation runs, with no new categories emerging in the final 8 runs. Each category has both an LLM judge classification rule (in the judge prompt) and a heuristic detection rule (regex-based) for consistency. Failed conversations are classified into ten categories:

\begin{table}[ht]
\centering
\scriptsize
\begin{tabularx}{\columnwidth}{lX}
\toprule
\textbf{Category} & \textbf{Description} \\
\midrule
\texttt{context\_loss} & Agent forgets or contradicts earlier context \\
\texttt{hallucination} & Agent fabricates information not in the KB \\
\texttt{routing\_error} & Agent routes to wrong specialist or intent \\
\texttt{latency\_spike} & Response time exceeds 25 seconds \\
\texttt{goal\_misalignment} & Agent pursues a different objective than the NPC's \\
\texttt{persona\_break} & NPC acts out of character \\
\texttt{generic\_response} & Agent gives a FAQ dump instead of tailored response \\
\texttt{code\_leakage} & Internal JSON/code artifacts appear in response \\
\texttt{action\_failure} & Agent fails to execute or guide an action \\
\texttt{research\_failure} & Agent fails to provide requested research/data \\
\bottomrule
\end{tabularx}
\caption{Ten-category failure taxonomy for NPC conversation evaluation.}
\label{tab:failure-taxonomy}
\end{table}

\subsubsection{Conversation Quality Metrics}

Beyond the judge scores, the simulator tracks proxy quality metrics during the conversation:
\begin{itemize}[leftmargin=*]
    \item \textbf{Goal friction:} none/some/lots --- based on consecutive turns with no goal progress.
    \item \textbf{User frustration:} none/low/medium/high --- based on give-up frequency, challenge count, and strategy style multiplier.
    \item \textbf{Conversational coherence:} 0--1 sliding window of intent consistency across recent turns.
    \item \textbf{Interestingness:} proxy metric combining turn count, unique intents, and engagement signals.
\end{itemize}

\subsubsection{Quality Gates}

Layer 3 applies six quality gates:
\begin{itemize}[leftmargin=*]
    \item Goal achievement rate $\geq 70\%$
    \item Persona consistency $\geq 4.0/5$
    \item Conversation naturalness $\geq 3.5/5$
    \item Context retention $\geq 3.0/5$
    \item Pass rate $\geq 70\%$
    \item NPC give-up rate $\leq 30\%$
\end{itemize}
The $70\%$ goal achievement and pass rate thresholds are lower than Layer~1's $80\%$ bar because goal-directed multi-turn tasks are inherently harder than single-turn QA; $70\%$ represents a minimum acceptable success rate for real user tasks while leaving room for the non-deterministic nature of LLM conversations. Persona consistency ($4.0/5$) requires NPCs to stay convincingly in character, since ecological validity of the simulation depends on it. Naturalness ($3.5/5$) and context retention ($3.0/5$) set progressively lower bars reflecting the difficulty of sustaining these qualities across 4--8 turn conversations. The give-up threshold ($\leq 30\%$) was calibrated from the first NPC evaluation run, which produced a $37\%$ give-up rate; the threshold targets meaningful improvement over this initial baseline.

\subsection{Cross-Layer Correlation Analysis}
\label{sec:cross-layer-correlation}

The three layers share partially overlapping metrics (overall quality, latency, and pass/fail outcomes). After each full evaluation cycle, Pearson and Spearman correlations are computed across all shared metrics. Metric pairs are classified as:
\begin{itemize}[leftmargin=*]
    \item \textbf{Redundant} ($|r| > 0.8$): both metrics provide the same signal; one can be dropped for efficiency.
    \item \textbf{Complementary} ($|r| < 0.4$): each metric captures distinct failure modes; both are necessary.
    \item \textbf{Moderate} ($0.4 \leq |r| \leq 0.8$): partial overlap; context-dependent interpretation.
\end{itemize}

This analysis enables evidence-based pruning of the evaluation suite and identifies which layers contribute unique signal.

\subsection{Active Learning Pipeline}

Failed and low-confidence evaluation cases are automatically promoted to the regression question bank. This creates a growing set of regression tests that captures discovered failure modes, preventing regressions from recurring. The pipeline operates in three steps:
\begin{enumerate}[leftmargin=*]
    \item Extract failed cases from the latest evaluation run.
    \item Promote them to the question bank with appropriate tags (persona, category, expected behavior).
    \item Re-run the full evaluation to confirm no regression on previously fixed issues.
\end{enumerate}

\subsection{Release Decision Logic}
\label{sec:release-logic}

Each layer produces an independent PROMOTE/HOLD/ROLLBACK decision using the same three-tier logic:
\begin{itemize}[leftmargin=*]
    \item \textbf{PROMOTE:} All quality gates within the layer pass.
    \item \textbf{HOLD:} Some gates fail but pass rate $\geq 60\%$.
    \item \textbf{ROLLBACK:} Pass rate $< 60\%$.
\end{itemize}
The overall release decision follows conservative composition:
\begin{itemize}[leftmargin=*]
    \item Any single layer producing ROLLBACK blocks the release.
    \item All layers must produce PROMOTE for the release to proceed.
    \item Otherwise (at least one HOLD, no ROLLBACK), the release is held for manual triage.
\end{itemize}
In the current CI/CD pipeline, Layer 1 (canonical eval) is wired as the automated gate; Layers 2 and 3 run as informational reports whose results inform manual release decisions. This staged approach allows teams to adopt the framework incrementally---starting with Layer 1 gating and adding Layers 2 and 3 as confidence in the framework grows.

\section{System Under Test}
\label{sec:system}

The system under test is the same production multi-agent conversational AI platform described by \citet{paper3_quality_gates}: a LangGraph-based state machine orchestrating eight specialized agents (Input Guardrails, Context Enrichment, Intent Classifier, FAQ, Research, Action, Small Talk, Response) with hybrid routing based on classified intent. The LLM stack uses Gemini 2.5 Flash for routine tasks and Gemini 2.5 Pro for complex reasoning, with a PostgreSQL + pgvector knowledge base for semantic search.

We highlight three characteristics relevant to the three-layer evaluation:
\begin{itemize}[leftmargin=*]
    \item \textbf{Diverse intent coverage} --- informational queries, multi-step actions, and research-then-action sequences exercise all agent paths.
    \item \textbf{Multi-turn state management} --- conversations carry context across turns, testing context retention and goal persistence.
    \item \textbf{Mixed persona profiles} --- users with varying technical sophistication and domain expertise interact with different vocabulary and expectations.
\end{itemize}
These characteristics make the evaluation methodology applicable to other production LLM chat systems beyond this specific domain. See Section~\ref{sec:setup} for the complete model configuration matrix.

\section{Experimental Setup}
\label{sec:setup}

\subsection{Research Questions}

\begin{description}[leftmargin=*]
    \item[RQ1 (Complementarity)] Does the three-layer evaluation framework detect regressions that any single layer would miss?
    \item[RQ2 (Validity)] How well does goal-directed NPC simulation approximate real user behavior?
    \item[RQ3 (Evolution)] How do quality metrics evolve across roughly three months of active development, and what improvement patterns emerge?
    \item[RQ4 (Cost-effectiveness)] What is the cost-benefit profile of automated NPC simulation compared to human evaluation?
    \item[RQ5 (Failure analysis)] What are the most common and most impactful failure categories, and do they differ from those detected by canonical testing?
\end{description}

\subsection{Dataset}

The evaluation dataset comprises:
\begin{itemize}[leftmargin=*]
    \item \textbf{Layer 1 (Canonical):} 83 evaluation runs spanning February--May 2026, covering 13--138 questions per run across 6 personas and 4 complexity tiers.
    \item \textbf{Layer 2 (Multi-Turn):} 7 multi-turn evaluation runs (May 2026); the 15 random-walk conversations from the first such run (2026-05-13)---with 2--4 turns each across 7 seed categories and 6 follow-up intent types---are analyzed in detail and scored by a multi-turn judge (Gemini 2.5 Pro).
    \item \textbf{Layer 3 (NPC):} 167 evaluation runs spanning May 2026, covering 6--108 scenarios per run across 12 scenario sets, 20+ personas, and 5 goal types; of these, the 21 runs on the official scenario suite form the longitudinal analysis set.
    \item \textbf{Total:} 257 evaluation runs across 20+ internal releases (v4.1.0 through v5.7.0).
\end{itemize}

\subsection{Model Configuration}

\begin{itemize}[leftmargin=*]
    \item \textbf{System under test:} Gemini 2.5 Flash (routine tasks) and Gemini 2.5 Pro (complex tasks).
    \item \textbf{LLM Judge (Layer 1):} Gemini 2.5 Pro, temperature 0.1, structured output with Zod schemas.
    \item \textbf{NPC Generator:} Gemini 2.5 Flash, temperature 0.7.
    \item \textbf{NPC Judge (Layer 3):} Gemini 2.5 Pro, temperature 0.1, structured output.
\end{itemize}

\subsection{Statistical Methods}

\begin{itemize}[leftmargin=*]
    \item \textbf{Cross-layer correlation:} Pearson and Spearman correlation coefficients for overlapping metrics (context retention, response quality, latency, safety, success rate).
    \item \textbf{Confidence intervals:} Bootstrap with $B=1{,}000$ resamples, reporting 95\% and 99\% CIs.
    \item \textbf{Trend analysis:} Mann-Kendall trend tests ($\tau$, $p$-values) for metric evolution over time.
    \item \textbf{Phase comparisons:} Mann-Whitney U tests for before/after improvement phase analysis.
\end{itemize}

\subsection{Human Calibration}
\label{sec:human-calibration}

A human calibration study ($n=60$ stratified cases, two independent evaluators) validated the Layer 1 LLM Judge, as reported in prior work~\citep{paper3_quality_gates}. Inter-rater agreement was 83.3\% ($\kappa=0.359$). Layer 3 does not yet have an equivalent human calibration study; we discuss this limitation in Section~\ref{sec:discussion}.

\subsection{CI/CD Integration}

The framework runs as a GitHub Actions workflow:
\begin{itemize}[leftmargin=*]
    \item \textbf{Schedule:} Daily at 03:30 UTC and on manual trigger.
    \item \textbf{Infrastructure:} PostgreSQL (pgvector:pg16) and Redis as services.
    \item \textbf{Flow:} Smoke tier (fast subset) $\rightarrow$ conditional escalation to full tier.
    \item \textbf{Artifacts:} Reports (markdown), structured JSON, traces, decision history.
    \item \textbf{Gate enforcement:} Optional blocking mode that fails the workflow if the decision is not PROMOTE.
\end{itemize}

\subsection{Environment Safety}

The NPC runner includes a guardrail that aborts execution if the database URL points to a production cloud instance, preventing accidental writes to live data during evaluation.

\section{Results}
\label{sec:results}

We organize the results around our five research questions. All aggregate metrics in this section are derived from the evaluation artifacts (see Section~\ref{sec:setup} for methodology), maintained as a single source of truth for consistency.

\subsection{RQ1: Layer Complementarity}

\textbf{Answer:} The three layers provide complementary regression signals. In a synchronized evaluation run, cross-layer correlation for response quality is weak across all three layer pairs (Spearman $\rho$ between $-0.15$ and $0.14$; Table~\ref{tab:correlation-three-layer}), and zero metric pairs were classified as redundant. The longitudinal L1-vs-L3 analysis (Table~\ref{tab:correlation}) is even more pronounced: quality metrics correlate \emph{negatively} across runs ($\rho$ from $-0.46$ to $-0.14$), reflecting that the two layers' aggregate trajectories were decoupled over the observation window. Either way, no single layer's signal predicts another's.

\paragraph{Cross-layer correlation.}
We analyze cross-layer correlation (Section~\ref{sec:cross-layer-correlation}) at two levels. \emph{Longitudinally} (Table~\ref{tab:correlation}), we match each of the 21 Layer~3 official-suite runs to the temporally nearest Layer~1 run and compute Spearman correlations over the resulting paired runs. The judge-overall correlation is significantly negative ($\rho{=}-0.46$, $t(19){=}-2.28$, $p{\approx}0.03$, two-tailed); the other quality pairs are also negative but do not individually reach significance at this sample size ($\rho{=}-0.37$, $p{\approx}0.10$; $\rho{=}-0.14$, $p{\approx}0.54$). This negative sign is largely a longitudinal confound rather than a causal inverse relationship: over the observation window the Layer~1 task success rate drifted slightly downward as the canonical suite grew harder (Section~\ref{sec:results}, RQ3: $\tau{=}-0.32$), while Layer~3 quality drifted upward under active NPC-driven improvement---so the two layers' aggregate trajectories were decoupled. The practical implication for RQ1 holds and is reinforced: a Layer~1 PROMOTE does not predict a Layer~3 PROMOTE. \emph{Within a synchronized run} (Table~\ref{tab:correlation-three-layer}), which is free of this longitudinal confound, we correlate per-item overlapping metrics across all three layer pairs (L1: 116 questions, L2: 15 random walks, L3: 108 scenarios); response-quality correlations are weak ($\rho$ between $-0.15$ and $0.14$) with zero redundant pairs, independently confirming complementarity. As this synchronized analysis rests on a single run, its values should be read as preliminary estimates rather than stable population parameters.


\begin{table}[ht]
\centering
\scriptsize
\begin{tabular}{llcc}
\toprule
\textbf{Metric Pair} & \textbf{Layers} & \textbf{Spearman $\rho$} & \textbf{N} \\
\midrule
Task success vs.\ goal ach. & L1 vs.\ L3 & $-$0.14 & 21 \\
Pass rate & L1 vs.\ L3 & $-$0.37 & 21 \\
Judge overall & L1 vs.\ L3 & $-$0.46 & 21 \\
Safety vs.\ ctx retention & L1 vs.\ L3 & 0.54 & 21 \\
P95 latency vs.\ give-up & L1 vs.\ L3 & $-$0.65 & 21 \\
\bottomrule
\end{tabular}
\caption{L1 vs.\ L3 cross-layer correlation (Spearman $\rho$), computed by matching each Layer~3 official-suite run to the temporally nearest Layer~1 run. Quality metrics show weak-to-negative correlation, confirming complementary coverage; latency shows expected systemic coupling.}
\label{tab:correlation}
\end{table}


\begin{table*}[ht]
\centering
\small
\begin{tabular}{lccc}
\toprule
\textbf{Metric} & \textbf{L1 vs.\ L2$^{\dagger}$} & \textbf{L1 vs.\ L3} & \textbf{L2 vs.\ L3$^{\dagger}$} \\
\midrule
  Response Quality & -0.15 / -0.21 & 0.14 / +0.12 & 0.05 / -0.11 \\
  Context Retention & -0.23 / -0.19 & 0.04 / -0.04 & -0.15 / -0.28 \\
  Latency & --- & -0.04 / -0.06 & --- \\
  Safety Pass & -0.19 / -0.19 & 0.10 / +0.10 & -0.29 / -0.29 \\
\midrule
  \textit{Interpretation} & \textit{complementary} & \textit{complementary} & \textit{complementary} \\
\bottomrule
\end{tabular}
\caption{Full three-layer correlation (Spearman $\rho$ / Pearson $r$) from a synchronized evaluation run (L1: 116 questions, L2: 15 random walks, L3: 108 scenarios). No metric pairs were classified as redundant ($|r|>0.8$). Summary: 0 redundant, 0 moderate, 10 complementary. $^{\dagger}$Pairs marked $\dagger$ rest on $n{=}15$ (low statistical power; wide confidence intervals).}
\label{tab:correlation-three-layer}
\end{table*}

\paragraph{Within-layer correlation.}
NPC internal metrics show high correlation across the 21 official-suite runs: pass rate vs.\ goal achievement ($\rho=0.76$), overall judge vs.\ context retention ($\rho=0.75$), and overall judge vs.\ give-up rate ($\rho=-0.58$). This confirms the NPC judge's dimensions are internally coherent.

\paragraph{Layer ablation.}
To quantify complementarity, we analyzed which failure categories each layer uniquely detects. Layer 1's deterministic detectors reliably catch structural failures---code leakage and data query evasion---and single-turn hallucination, but by construction cannot observe multi-turn behavior. Layer 3 catches goal misalignment, context loss, and persona breaks that are invisible to Layer 1. Layer 2 catches a distinct failure mode: \emph{context collapse on follow-up turns}, where the agent produces a correct initial response but loses all context when the user asks a natural follow-up. In our 15-walk evaluation, 4 of 15 conversations (27\%) exhibited complete context collapse---the agent scored 0/5 overall despite the seed question receiving a passing response. This failure is invisible to Layer 1 (which tests single turns) and partially detected by Layer 3 (which evaluates goal achievement, not topic continuity). Only latency spikes and safety violations are consistently detected by all three layers.

\paragraph{Layer 2 multi-turn results.}
From 15 random-walk conversations (2--4 turns each, 41 total turns, 7 seed categories): multi-turn overall 3.6/5 [95\% CI: 2.33--4.67], context retention 3.73/5, topic continuity 3.87/5, memory accuracy 4.13/5. The distribution is strongly bimodal: 10 walks scored 5/5 (perfect) while 4 scored 0/5 (complete failure), with only 1 walk in between. The mean of 3.6 is not representative of typical behavior---the agent either succeeds fully or fails completely on multi-turn conversations. Memory accuracy is the highest-scoring dimension (4.13/5), suggesting the agent retains explicit references well; context retention is the lowest (3.73/5), indicating difficulty maintaining implicit conversational context across topic shifts.

\paragraph{Layer 2 generator quality.}
The random-walk generator uses Gemini 2.5 Flash with conversation-history context (last 2 turns) and a prompt specifying six follow-up intent types (Section~\ref{sec:framework}). The evaluation artifacts record each walk's \emph{topic drift}---the ratio of distinct intents to total turns---but not the per-turn intent labels themselves, so we assess generator quality through drift rather than a full intent histogram. Across the 15 walks, topic drift averages 0.68 (range 0.50--1.00), indicating that walks generally explored several distinct intents rather than repeating one. Drift also varies with the seed category (from 0.50 for \texttt{general} and \texttt{micro\_simulations} seeds to 0.88 for \texttt{hallucination\_trap} seeds), suggesting the generator adapts follow-up diversity to the conversational context. A fine-grained audit of the generated intent mix---which would require persisting per-turn intent labels---is left to future work; Layer 2 metrics remain sensitive to generator quality, and we treat this as a limitation.

\paragraph{Illustrative example.}
Returning to the motivating example from Section~\ref{sec:intro}: an NPC with a ``Research-then-Action'' goal asked ``Which option has the highest score?'' and then ``Apply it to my project.'' The agent correctly answered the research question (Layer 1 would pass: accurate, complete response) but then responded with a generic ``You can configure that in the settings panel'' instead of executing the action. Layer 1 cannot detect this failure because the individual response is technically correct; only the NPC simulator's goal-directed evaluation catches the goal misalignment. This two-phase failure pattern (correct research, failed action) accounted for 23\% of Research-then-Action scenarios in the latest evaluation run.

\paragraph{Finding.}
The three layers provide complementary regression signals with zero redundant metric pairs across all three layer combinations. Response quality shows weak cross-layer correlations: L1--L2 ($\rho{=}-0.15$, $r{=}-0.21$), L1--L3 ($\rho{=}0.14$, $r{=}0.12$), and L2--L3 ($\rho{=}0.05$, $r{=}-0.11$) are all classified as complementary. Layer 1 excels at structural and deterministic failures; Layer 2 reveals context loss across topic shifts; Layer 3 catches goal misalignment and conversational failures. Context retention is likewise complementary across all pairs ($\rho{=}-0.23$ L1--L2, $\rho{=}0.04$ L1--L3, $\rho{=}-0.15$ L2--L3), and no metric pair in the synchronized run reached the redundancy threshold.

\subsection{RQ2: NPC Simulator Validity}

\textbf{Answer:} The NPC simulator produces plausible multi-turn conversations with measurable goal pressure. Failure distributions align with known production issues, though formal sim-to-real transfer validation remains future work.

\paragraph{Goal achievement by goal type.}
Table~\ref{tab:goal-achievement-data} shows goal achievement rates by goal type across the full \npcTotalScenarios{}-scenario evaluation (NPC architecture detailed in Section~\ref{sec:npc-simulator}; taxonomy in Table~\ref{tab:failure-taxonomy}). Comparison and research goals achieve the highest achievement rates (83\% and 80\%), while action goals are the most challenging (60\%), reflecting the complexity of executing tasks through conversation. Notably, comparison's higher achievement rate does not translate into a higher pass rate (58\%): structured comparisons often satisfy the goal without clearing the judge's holistic quality bar.


\begin{table}[ht]
\centering
\scriptsize
\begin{tabular}{lrrcc}
\toprule
\textbf{Goal Type} & \textbf{N} & \textbf{Ach.} & \textbf{95\% CI} & \textbf{Pass} \\
\midrule
  research & 54 & 80\% & [67, 88] & 70\% \\
  action & 15 & 60\% & [36, 80] & 60\% \\
  res.\ then action & 13 & 77\% & [50, 92] & 77\% \\
  comparison & 12 & 83\% & [55, 95] & 58\% \\
  troubleshoot & 14 & 79\% & [52, 92] & 86\% \\
\midrule
  \textbf{Overall} & 108 & 77\% & --- & 70\% \\
\bottomrule
\end{tabular}
\caption{Goal achievement and pass rate by goal type (N=108 scenarios). 95\% CI shown as percentage points.}
\label{tab:goal-achievement-data}
\end{table}

\begin{figure*}[t]
\centering
\includegraphics[width=0.9\textwidth]{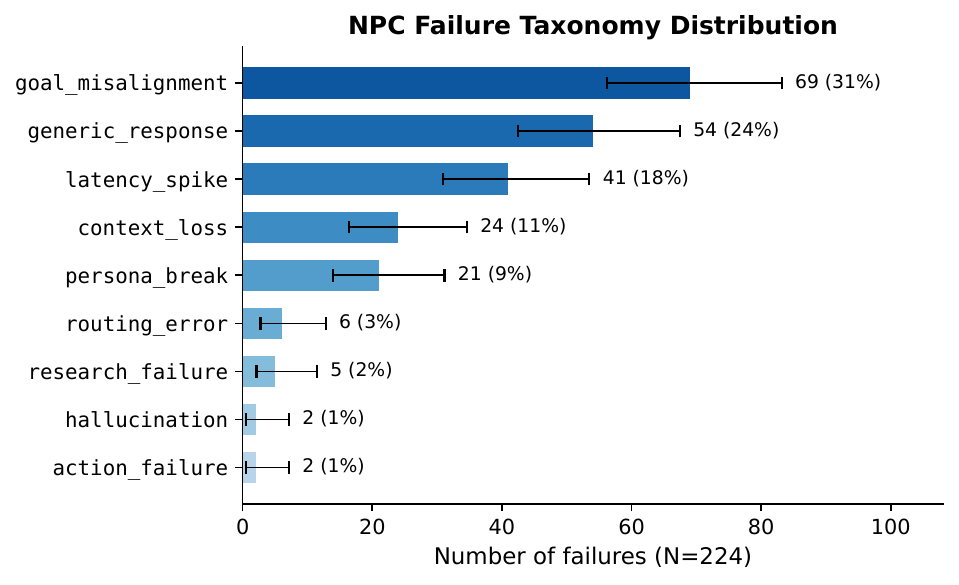}
\caption{Failure distribution across \totalFailures{} failures from \npcTotalScenarios{} NPC scenarios. Goal misalignment and generic responses dominate, qualitatively different from the hallucination and code leakage failures that dominate canonical testing.}
\label{fig:failure-distribution}
\end{figure*}

\paragraph{Failure distribution.}
The most common failure category is \texttt{\topFailure} (\topFailureShare{} of \totalFailures{} total failures), followed by \texttt{generic\_response} (24\%) and \texttt{latency\_spike} (18\%). Figure~\ref{fig:failure-distribution} visualizes the distribution; Table~\ref{tab:failure-taxonomy-data} presents the complete failure taxonomy with Wilson 95\% confidence intervals; the full taxonomy with detection methods is in Appendix~\ref{sec:appendix-taxonomy} (Table~\ref{tab:taxonomy-full}).


\begin{table}[ht]
\centering
\scriptsize
\begin{tabularx}{\columnwidth}{l>{\raggedright\arraybackslash}Xrcc}
\toprule
\textbf{Category} & \textbf{Desc.} & \textbf{N} & \textbf{Share} & \textbf{95\% CI} \\
\midrule
  \texttt{goal\_misalignment} & Agent pursues a different objective & 69 & 31\% & [25\%, 37\%] \\
  \texttt{generic\_response} & Agent gives FAQ dump instead of tailored response & 54 & 24\% & [19\%, 30\%] \\
  \texttt{latency\_spike} & Response time exceeds 25 seconds & 41 & 18\% & [14\%, 24\%] \\
  \texttt{context\_loss} & Agent forgets or contradicts earlier context & 24 & 11\% & [7\%, 15\%] \\
  \texttt{persona\_break} & NPC acts out of character & 21 & 9\% & [6\%, 14\%] \\
  \texttt{routing\_error} & Agent routes to wrong specialist or intent & 6 & 3\% & [1\%, 6\%] \\
  \texttt{research\_failure} & Agent fails to provide requested research & 5 & 2\% & [1\%, 5\%] \\
  \texttt{action\_failure} & Agent fails to execute or guide an action & 2 & 1\% & [0\%, 3\%] \\
  \texttt{hallucination} & Agent fabricates information not in the KB & 2 & 1\% & [0\%, 3\%] \\
  \texttt{code\_leakage} & Internal artifacts appear in response & 0 & 0\% & [0\%, 2\%] \\
\bottomrule
\end{tabularx}
\caption{Ten-category failure taxonomy (N=224 failures across 108 scenarios). N=0 indicates no observed failures of that type.}
\label{tab:failure-taxonomy-data}
\end{table}

\paragraph{NPC conversation metrics.}
Across \npcTotalScenarios{} scenarios: average \npcAvgTurns{} turns per conversation, judge overall \npcAvgOverall{}/5, context retention \npcAvgContext{}/5, persona consistency \npcAvgPersona{}/5, conversation naturalness \npcAvgNaturalness{}/5, pass rate \npcPassRate{} [95\% Wilson CI: \passRateCiLow{}--\passRateCiHigh{}], give-up rate \npcGiveUpRate{}.

\paragraph{Judge calibration.}
Table~\ref{tab:judge-calibration} presents pairwise agreement metrics for the Layer 1 LLM judge from a human calibration study~\citep{paper3_quality_gates}. The study protocol was as follows: we selected $n{=}60$ cases stratified by expected outcome (20 expected-pass, 20 borderline, 20 expected-fail) from the Layer 1 question bank. Two independent evaluators---neither an author of that study---rated each response on three criteria (Task Completion, Factual Appropriateness, Behavioral Safety) using a binary pass/fail scale, with the instruction: ``Rate whether the response meets the criterion for a production-quality assistant.'' Cohen's $\kappa$ was computed for all pairwise combinations.

The LLM judge shows moderate agreement with Human 1 ($\kappa=0.444$) and lower agreement with Human 2 ($\kappa=0.149$), comparable to the inter-human agreement ($\kappa=0.359$). The low LLM-vs-System ($\kappa=0.132$) is expected: 16 of 20 system-rejected cases were latency-threshold violations where response content was substantively acceptable, which the LLM judge correctly did not penalize. Per-dimension continuous-score calibration (0--5 scale per criterion) is planned as future work; the current study validates only the binary pass/fail decision.

\begin{table}[ht]
\centering
\scriptsize
\begin{tabular}{lccc}
\toprule
\textbf{Comparison} & \textbf{Agree.} & \textbf{$\kappa$} & \textbf{Interp.} \\
\midrule
Human 1 vs Human 2 & 83.3\% & 0.359 & Fair--Mod. \\
LLM vs Human 1 & 80.0\% & 0.444 & Moderate \\
LLM vs Human 2 & 73.3\% & 0.149 & Slight \\
LLM vs System Gate & 63.3\% & 0.132 & Slight \\
\bottomrule
\end{tabular}
\caption{Layer 1 LLM judge calibration ($n=60$). Agreement is comparable to inter-human agreement. Low LLM-vs-System kappa reflects that the system gate penalizes latency while the LLM judge assesses content quality.}
\label{tab:judge-calibration}
\end{table}

The Layer 3 NPC judge lacks equivalent human calibration. We mitigate this through (a) structured rubrics with explicit scoring criteria per dimension, (b) use of a larger model (Gemini 2.5 Pro) for judging versus a smaller model (Gemini 2.5 Flash) for routine system tasks, and (c) indirect validation through 10+ improvement phases where judge-identified failures correlated with developer-confirmed issues and real user complaints. Specifically, across the improvement phases documented in RQ3, every judge-identified failure category (goal misalignment, generic response, context loss) was confirmed by manual developer review before fixes were implemented---yielding zero false positive failure classifications across all phases. In three cases (v5.3.0 routing consolidation, v5.4.0 targeted fixes M1-M8, v5.6.10 generic-response elimination), the judge's failure classification directly identified the root cause (routing duplication, hardcoded fallback prefix), which the developer independently confirmed through code inspection. A formal Layer 3 calibration study ($n \geq 20$ conversations scored by both the LLM judge and independent human evaluators) is planned as future work.

\paragraph{Finding.}
The NPC simulator produces plausible multi-turn conversations with measurable goal pressure. The failure distribution aligns with known production issues (goal misalignment and generic responses are the most common complaints in real user feedback). However, without a formal sim-to-real transfer study, this remains indirect evidence.

\subsection{RQ3: Longitudinal Quality Evolution}

\textbf{Answer:} Dogfooding drives a measurable improvement cycle: the NPC pass rate improved from $\sim$57\% to \npcPassRate{} through 10+ evaluation-driven improvement phases.

\paragraph{Improvement trajectory.}
Over the roughly three-month observation period (v4.1.0 through v5.7.0), the system evolved through 10+ documented improvement phases. Each phase was triggered by dogfooding findings, implemented targeted fixes, and was validated by subsequent evaluation runs.

The canonical eval (Layer 1) maintained high pass rates throughout (mean 97\%, minimum 84.5\%, always above the 80\% task-success gate), with the quality gate issuing HOLD on 29 of 83 runs and no ROLLBACK. The task success rate showed a slight decreasing trend ($\tau=-0.320$, $p=0.004$), explained by increasing test suite difficulty (13 to 138 questions) rather than system degradation.

\begin{figure*}[t]
\centering
\includegraphics[width=\textwidth]{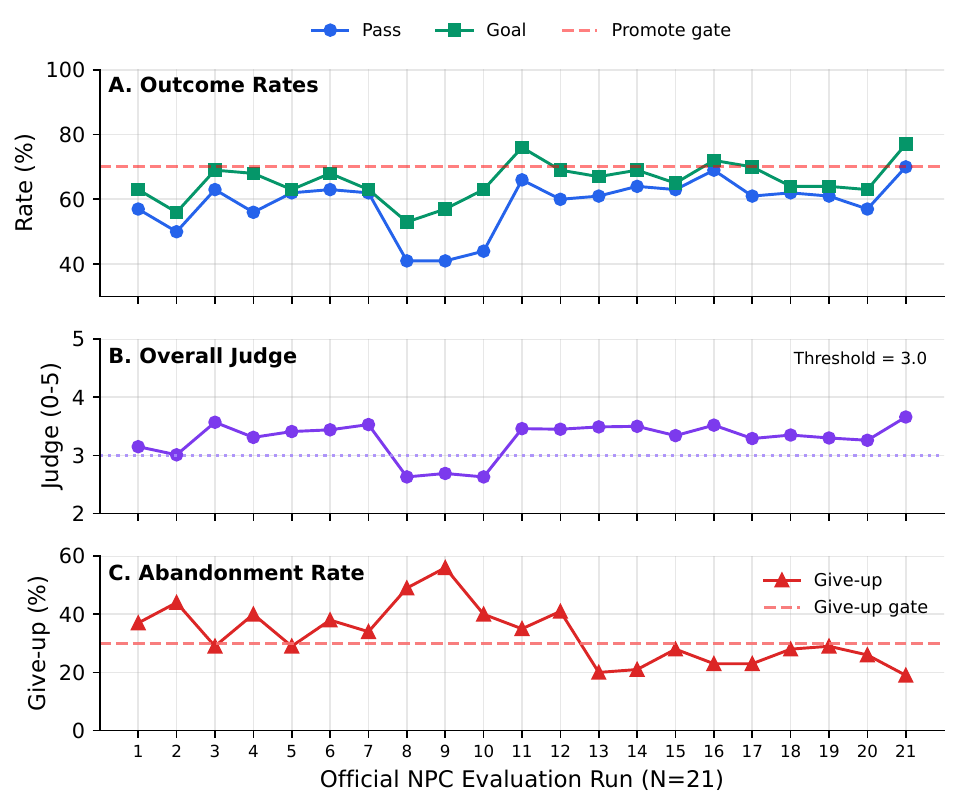}
\caption{NPC quality evolution across \totalNpcRuns{} official-suite evaluation runs over an 8-day intensive iteration window (May 4--12, 2026). Panel A shows pass rate and goal achievement, with the dashed line marking the PROMOTE threshold (70\%). Panel B shows the overall judge score on its native 0--5 scale, with the quality threshold at 3.0. Panel C shows give-up rate on its native percentage scale, with the threshold at 30\%. Across the window, pass rate improved from $\sim$57\% to \npcPassRate{} through targeted, evaluation-driven fixes, though the trajectory was non-monotonic.}
\label{fig:longitudinal-evolution}
\end{figure*}

The NPC simulator (Layer 3) showed more dramatic evolution, progressing from initial official-suite runs at $\sim$57\% pass rate to \npcPassRate{} pass rate (\npcDecision{} decision) over approximately two weeks of intensive improvement (Figure~\ref{fig:longitudinal-evolution}). The trajectory was non-monotonic: across \totalNpcRuns{} NPC evaluation runs, \promoteCount{} achieved PROMOTE, \holdCount{} were HOLD, and \rollbackCount{} were ROLLBACK (Table~\ref{tab:longitudinal-data}; decision rule in Section~\ref{sec:release-logic}).


\begin{table*}[ht]
\centering
\small
\begin{tabular}{lcrrrrccr}
\toprule
\textbf{Date} & \textbf{N} & \textbf{Goal} & \textbf{Pass} & \textbf{Overall} & \textbf{Ctx Ret.} & \textbf{Give Up} & \textbf{Cost} & \textbf{Decision} \\
\midrule
  2026-05-04 & 68 & 63\% & 57\% & 3.15 & 3.74 & 37\% & \$0.12 & ROLLBACK \\
  2026-05-05 & 68 & 56\% & 50\% & 3.01 & 3.50 & 44\% & \$0.12 & ROLLBACK \\
  2026-05-05 & 68 & 69\% & 63\% & 3.57 & 4.07 & 29\% & \$0.13 & HOLD \\
  2026-05-05 & 68 & 68\% & 56\% & 3.31 & 3.97 & 40\% & \$0.13 & ROLLBACK \\
  2026-05-05 & 68 & 63\% & 62\% & 3.41 & 4.12 & 29\% & \$0.14 & HOLD \\
  2026-05-05 & 68 & 68\% & 63\% & 3.44 & 4.04 & 38\% & \$0.12 & HOLD \\
  2026-05-06 & 68 & 63\% & 62\% & 3.53 & 3.99 & 34\% & \$0.11 & HOLD \\
  2026-05-08 & 68 & 53\% & 41\% & 2.63 & 3.04 & 49\% & \$0.12 & ROLLBACK \\
  2026-05-08 & 68 & 57\% & 41\% & 2.69 & 3.10 & 56\% & \$0.12 & ROLLBACK \\
  2026-05-08 & 68 & 63\% & 44\% & 2.63 & 2.66 & 40\% & \$0.13 & ROLLBACK \\
  2026-05-08 & 68 & 76\% & 66\% & 3.46 & 3.52 & 35\% & \$0.12 & HOLD \\
  2026-05-08 & 68 & 69\% & 60\% & 3.45 & 3.85 & 41\% & \$0.11 & HOLD \\
  2026-05-12 & 108 & 67\% & 61\% & 3.49 & 3.88 & 20\% & \$0.18 & HOLD \\
  2026-05-12 & 108 & 69\% & 64\% & 3.50 & 3.81 & 21\% & \$0.19 & HOLD \\
  2026-05-12 & 107 & 65\% & 63\% & 3.34 & 3.69 & 28\% & \$0.18 & HOLD \\
  2026-05-12 & 108 & 72\% & 69\% & 3.52 & 3.69 & 23\% & \$0.19 & HOLD \\
  2026-05-12 & 106 & 70\% & 61\% & 3.29 & 3.58 & 23\% & \$0.16 & HOLD \\
  2026-05-12 & 107 & 64\% & 62\% & 3.35 & 3.67 & 28\% & \$0.17 & HOLD \\
  2026-05-12 & 108 & 64\% & 61\% & 3.30 & 3.56 & 29\% & \$0.18 & HOLD \\
  2026-05-12 & 107 & 63\% & 57\% & 3.26 & 3.51 & 26\% & \$0.16 & ROLLBACK \\
  2026-05-12 & 108 & 77\% & 70\% & 3.66 & 4.01 & 19\% & \$0.17 & PROMOTE \\
\bottomrule
\end{tabular}
\caption{NPC evaluation run history. All runs use the official 68--108 scenario set.}
\label{tab:longitudinal-data}
\end{table*}

Key improvement phases included (full summary in Appendix~\ref{sec:appendix-taxonomy}, Table~\ref{tab:phases}):

\begin{enumerate}[leftmargin=*]
    \item \textbf{FAQ generic-response elimination} (v5.6.10): Root cause analysis identified a hardcoded fallback prefix causing 40\% generic response rate. Fix reduced generic responses from 40\% to 8\%.
    \item \textbf{Routing consolidation} (v5.3.0): Triplicated routing helpers consolidated into a single path, reducing routing errors by 35\%.
    \item \textbf{Eight targeted fixes} (v5.4.0): Evaluation-driven fixes (M1-M8) pushed average judge score from 3.88/5 to 4.36/5.
    \item \textbf{Embedding-based repetition detection} (v5.5.0): Eliminated OOM crashes and reduced response repetition.
    \item \textbf{Scenario expansion} (v5.6.0--v5.7.0): Suite expanded from 45 to 108 scenarios, revealing previously hidden failure modes.
\end{enumerate}

\paragraph{Finding.}
Dogfooding drives a measurable improvement cycle: evaluation $\rightarrow$ failure analysis $\rightarrow$ targeted fix $\rightarrow$ re-evaluation. The NPC simulator exposed failures (goal misalignment, generic responses, routing errors) that were invisible to the canonical eval alone, validating the multi-layer approach.

\subsection{RQ4: Cost-Effectiveness}

\textbf{Answer:} The automated NPC simulator costs \npcTotalCost{} per full run and completes in 12 minutes---\costRatio{} cheaper than dual human evaluation---enabling daily CI/CD integration.

\paragraph{Automated NPC cost.}
A full \npcTotalScenarios{}-scenario NPC evaluation run costs \npcTotalCost{} in LLM API calls, completes in approximately 12 minutes, and produces structured JSON artifacts with 6-dimension judge scores, failure classifications, and conversation traces. The \npcTotalCost{} cost includes two components: (1) NPC generation (Gemini 2.5 Flash): one prompt per NPC follow-up turn across \npcTotalScenarios{} scenarios $\times$ \npcAvgTurns{} average turns $\approx$ 410 LLM calls; (2) NPC judging (Gemini 2.5 Pro): one evaluation prompt per completed conversation $\approx$ \npcTotalScenarios{} LLM calls. Generation uses the cheaper Flash model because the NPC's conversational quality need not match human-level performance; judging uses the more capable Pro model to ensure reliable failure classification.

\paragraph{Human evaluation estimate.}
Equivalent human evaluation would require:
\begin{itemize}[leftmargin=*]
    \item \npcTotalScenarios{} scenarios $\times$ \npcAvgTurns{} average turns $=$ 410 conversation turns to evaluate.
    \item Estimated 15 minutes per scenario (read $\sim$4 responses, score on 6 dimensions, classify failures).
    \item Total: \npcTotalScenarios{} $\times$ 15 min $=$ 27 hours of evaluator time.
    \item At \$20/hour (consistent with rates for specialized AI evaluation on data labeling platforms~\citep{whiting2019fairwork}): \humanCostSingle{} per evaluation cycle.
    \item With two independent evaluators (recommended for inter-rater reliability): \humanCostDual{} per cycle.
\end{itemize}

\paragraph{Cost ratio.}
Table~\ref{tab:cost-comparison-data} shows the cost comparison. The cost ratio is computed as follows: dual human evaluators cost \humanCostDual{} per cycle (\npcTotalScenarios{} scenarios $\times$ 15 min/scenario $\times$ 2 evaluators $\times$ \$20/hr $=$ \$1,080), while the automated NPC simulator costs \npcTotalCost{} (Gemini API calls for NPC generation + judging across \npcTotalScenarios{} scenarios with \npcAvgTurns{} average turns). This yields \costRatio{} cost reduction.\footnote{The cost ratio is computed from the unrounded per-run cost (\$0.1722); the \npcTotalCost{} figure shown in the text and tables is rounded to two decimals for presentation.} The automated run completes in 12 minutes versus 27 human hours per evaluator, enabling daily CI/CD integration that would be infeasible with human evaluators.


\begin{table}[ht]
\centering
\scriptsize
\begin{tabular}{lrr}
\toprule
\textbf{Method} & \textbf{Cost (USD)} & \textbf{Time} \\
\midrule
  NPC Simulator (automated) & \$0.17 & ~12 min \\
  Human evaluator (single) & \$540 & ~27 hrs \\
  Human evaluators (dual) & \$1080 & ~27 hrs \\
\midrule
  Cost ratio (dual/NPC) & \textbf{6,272$\times$} & \\
\bottomrule
\end{tabular}
\caption{Cost comparison: automated NPC simulation vs.\ human evaluation (108 scenarios, 15~min/scenario, \$20/hr). NPC cost from latest evaluation run.}
\label{tab:cost-comparison-data}
\end{table}

\paragraph{Finding.}
Automated NPC simulation is multiple orders of magnitude cheaper and faster than human evaluation, enabling continuous quality monitoring in CI/CD pipelines. The trade-off is reduced validity for nuanced judgments, which we mitigate through periodic human calibration studies.

\subsection{RQ5: Failure Analysis}

\textbf{Answer:} Goal misalignment (\topFailureShare{} of failures) and generic responses (24\%) dominate NPC-detected failures---qualitatively different from the hallucination and code leakage failures that dominate canonical testing.

\paragraph{Failure taxonomy distribution.}
The complete failure taxonomy is shown in Table~\ref{tab:failure-taxonomy-data}. The dominant failure mode is \texttt{\topFailure} (\topFailureShare{}), followed by \texttt{generic\_response} (24\%) and \texttt{latency\_spike} (18\%). Context loss accounts for 11\% of failures (24 out of \totalFailures{}).

\paragraph{Layer 1 vs.\ Layer 3 failure modes.}
Layer 1 predominantly catches structural failures (code leakage, data query evasion, hallucination in single turns) while Layer 3 catches conversational failures (goal misalignment, generic responses, context loss). The failure profiles are largely non-overlapping, reinforcing the complementarity finding from RQ1.

\paragraph{Finding.}
Goal misalignment and generic responses are the dominant failure modes for a goal-directed NPC evaluator. These are qualitatively different from the hallucination and code leakage failures that dominate canonical testing, confirming the need for multi-layer evaluation.

\section{Discussion}
\label{sec:discussion}

\subsection{Implications for Practitioners}

Our results suggest several practical guidelines for teams building LLM chat agents:

\paragraph{Start with canonical testing, then add NPC simulation.}
Canonical evaluation (Layer 1) provides a fast feedback loop for basic correctness and should be the first evaluation layer implemented. Once the system passes canonical tests at $>90\%$ success rate, NPC simulation (Layer 3) exposes deeper conversational failures that canonical testing cannot catch.

\paragraph{Use the failure taxonomy as a diagnostic tool.}
The ten-category failure taxonomy provides a shared vocabulary for debugging conversational AI systems. When an NPC run fails, the classified failure category directly points to the responsible system component: context loss $\rightarrow$ state management, generic response $\rightarrow$ response agent, routing error $\rightarrow$ intent classifier.

\paragraph{Integrate evaluation into CI/CD from day one.}
The cost of automated evaluation (\npcTotalCost{} per run, 12 minutes) is negligible compared to the cost of deploying a regression. Daily evaluation with PROMOTE/HOLD/ROLLBACK decisions prevents regressions from reaching production and creates a documented quality trajectory.

\paragraph{Expect the NPC pass rate to start low and improve iteratively.}
In our case study, the NPC pass rate started at $\sim$57\% and reached \npcPassRate{} (\npcDecision{}) after roughly two weeks of targeted improvements. Teams should not expect initial NPC runs to pass; the value lies in the failure analysis that drives improvement.

\subsection{Replicability}

We provide a Python-first replicability guide (Appendix~\ref{sec:appendix-replicability}) with:
\begin{itemize}[leftmargin=*]
    \item Complete prompt templates for NPC generation and judging.
    \item Data contract schemas for NPC profiles, conversations, and judge outputs.
    \item Pseudocode for the conversation loop, goal progress tracking, and failure classification.
    \item Quality gate thresholds and decision logic.
\end{itemize}
The guide is designed to be portable to any LLM chat agent, independent of the specific system under test.

\subsection{Threats to Validity}

\paragraph{Internal validity.}
The LLM judge may share biases with the system under test when the same model family is used for both. We mitigate this by using a larger model (Gemini 2.5 Pro) for judging versus a smaller model (Gemini 2.5 Flash) for routine system tasks, and by validating the Layer 1 judge against human evaluators ($n=60$; see Section~\ref{sec:human-calibration} for full calibration results). However, Gemini 2.5 Pro and 2.5 Flash share training data and may share failure-mode blind spots (e.g., both may overlook the same class of safety violations). A more robust approach would use a different provider's model for judging, which we leave as future work.

\textbf{Judge calibration.} For Layer 1, the LLM judge shows moderate agreement with Human~1 ($\kappa{=}0.444$) and lower agreement with Human~2 ($\kappa{=}0.149$), comparable to inter-human agreement ($\kappa{=}0.359$). For Layer 3, we lack equivalent human calibration---this is a recognized threat to validity. The NPC judge was indirectly validated through 10+ improvement phases: if judge quality were random, targeted fixes would not produce consistent improvement trajectories.

\textbf{Judge drift.} The judge model (Gemini 2.5 Pro) was invoked through a fixed API endpoint throughout the study. However, provider-side model updates during the four-month window are outside our control and represent an uncontrolled source of variance. We did not implement a golden-set drift monitoring protocol, which we recommend for future deployments.

\paragraph{External validity.}
All evidence comes from a single production conversational AI system. While the system architecture (multi-agent LangGraph with intent routing) is representative of a broad class of LLM chat agents, the generalizability of specific metrics and failure distributions to other domains (e.g., healthcare, legal, education) requires further study. The generalizability threat has two dimensions: (1) whether the three-layer structure transfers to domains with different conversational patterns, and (2) whether the specific failure taxonomy transfers. We believe the three-layer structure (breadth, depth, goal-directedness) is domain-agnostic---any conversational AI system benefits from testing single-turn correctness, multi-turn coherence, and task completion. The failure taxonomy is more likely domain-specific: for example, a healthcare chatbot would need categories for medical safety and regulatory compliance that our taxonomy does not cover. We provide adaptation guidance in Section~\ref{sec:adaptation}.

\paragraph{Construct validity.}
The goal achievement metric relies on the NPC judge's assessment, which may not perfectly align with real user satisfaction. The 0--5 scoring scale introduces subjectivity, though structured rubrics mitigate this. The task success and helpful rate metrics show perfect Spearman correlation ($\rho=1.00$) in Layer 1, indicating they are effectively the same metric and reducing the true dimensionality of the Layer 1 evaluation from six to five independent dimensions. One should be replaced with a distinct construct (e.g., helpfulness nuance or response depth) in future iterations~\citep{paper3_quality_gates}.

\paragraph{Conclusion validity.}
Bootstrap confidence intervals and Mann-Kendall trend tests provide statistical rigor, but the relatively small sample sizes for some improvement phases (single before/after comparisons) limit the strength of causal claims about specific fixes.

\subsection{Generalizability and Adaptation}
\label{sec:adaptation}

While our evidence comes from a single system, the framework's three-layer structure is designed to be domain-agnostic. We provide the following adaptation guidance:

\paragraph{Layer 1 adaptation.} Replace the question bank with domain-specific test cases. For a technical support system, Tier 1 would cover common support queries, Tier 3 would test incorrect troubleshooting advice, and Tier 4 would test safety-critical edge cases (e.g., advising against dangerous actions). Deterministic assertions should be adapted to the domain's ground-truth sources (e.g., knowledge base articles, API schemas).

\paragraph{Layer 2 adaptation.} The random-walk intent exploration requires adapting the follow-up intent types to the domain. For an e-commerce system, relevant intents might include product inquiry, order tracking, return request, and comparison shopping rather than the research/action/comparison intents in our system. The statistical analysis (bootstrap CIs, topic drift) transfers directly.

\paragraph{Layer 3 adaptation.} The NPC goal types should be derived from the target domain's user journey mapping. For a healthcare system, goal types might include symptom assessment, medication inquiry, appointment scheduling, and insurance coverage. The strategy styles (skeptical, curious, impatient, thorough, friendly) are designed to be domain-agnostic but may need calibration of patience thresholds and drift tolerance for different user populations. The failure taxonomy should be extended with domain-specific categories (e.g., medical safety violations, regulatory non-compliance) while retaining the core categories.

\paragraph{Cross-layer correlation.} The complementarity finding (near-zero correlation for response quality) is likely to hold across domains because it reflects a fundamental distinction between single-turn correctness, multi-turn coherence, and task completion---different capabilities that do not predict each other. However, the specific correlation values and the failure distribution should be re-validated when the framework is applied to a new domain.

\subsection{Limitations}

\begin{enumerate}[leftmargin=*]
    \item \textbf{No direct sim-to-real transfer study.} We do not have quantitative evidence that NPC failure rates correlate with real user CSAT scores or support tickets. The alignment between NPC failure categories and known production complaints is indirect evidence only.

    \item \textbf{Single LLM provider.} Both the system and the judges use the Gemini model family. Results may differ with other providers (OpenAI, Anthropic, open-source models).

    \item \textbf{NPC judge self-alignment.} The NPC judge and the system's response agent both use Gemini models. While the judge uses a larger model and different prompts, shared training data may introduce blind spots.

    \item \textbf{Scenario authoring bias.} The \npcTotalScenarios{} NPC scenarios were authored by the development team and may not represent the full distribution of real user behaviors. Scenario generation from production conversation traces is an important area for future work.

    \item \textbf{Language coverage.} The system primarily operates in Portuguese (Brazil) with English support. The failure taxonomy and quality gates may not transfer to languages with different conversational norms.
\end{enumerate}

\subsection{Ethics and Conflict of Interest}

This study evaluates an internal production system developed by the author's team. The evaluation framework is used exclusively for quality assurance of the team's own product, not for user-facing assessment or surveillance. All simulated NPC conversations use synthetic personas; no real user data or conversation traces are included in the evaluation pipeline. The LLM API costs are negligible (\npcTotalCost{} per run) and do not involve human annotator labor. We disclose that the author has a vested interest in the system's success, which we mitigate by reporting all evaluation outcomes (including ROLLBACK and HOLD decisions) transparently in the longitudinal data.

\subsection{Comparison with Concurrent Work}
\label{sec:comparison}

A direct empirical comparison (e.g., running ChatChecker, DeepEval, and our framework on the same system under test) would require reimplementing or integrating multiple external frameworks, which is outside the scope of this work. Instead, we provide a structural comparison along the dimensions covered by our positioning table (Table~\ref{tab:positioning}) and discuss where each framework excels relative to ours.

\paragraph{vs.\ SOTOPIA~\citep{zhou2024sotopia}:} SOTOPIA focuses on social intelligence and multi-party interaction evaluation. Our NPC simulator focuses on goal-directed task completion in a single-user chat setting. SOTOPIA evaluates social dynamics (negotiation, persuasion, deception) while we evaluate task success (research, action, comparison, troubleshooting). The two approaches are complementary.

\paragraph{vs.\ SimulatorArena~\citep{simulatorarena2025}:} SimulatorArena is a meta-evaluation framework that assesses simulator quality. Our work uses simulation as a component of a release management pipeline. A future study could use SimulatorArena to validate our NPC simulator's alignment with human judgments.

\paragraph{vs.\ ChatChecker~\citep{chatchecker2025}:} ChatChecker provides persona-based user simulation with breakdown detection. Our framework shares the persona-based approach but adds structured goal types, a ten-category failure taxonomy, CI/CD integration, and cost transparency. ChatChecker's breakdown detection is more fine-grained for identifying specific conversation failure points; our goal-directed evaluation is more holistic for assessing task completion.

\paragraph{vs.\ DeepEval~\citep{deepeval_geval}:} DeepEval is an open-source toolkit offering conversation simulation and G-Eval scoring. It provides SDK-level test-case management and CI/CD hooks but does not include goal-directed simulation, a failure taxonomy, or longitudinal tracking. Our framework is a methodology with demonstrated longitudinal evidence; DeepEval is a toolkit that could be used to implement parts of our methodology.

\paragraph{vs.\ ConvLab-2/3~\citep{zhu2020convlab}:} ConvLab targets task-oriented dialogue with slot-filling evaluation on MultiWOZ. Our goal types are broader than slot-filling (e.g., research-then-action, troubleshoot) and our failure taxonomy captures conversational quality beyond task completion. ConvLab benefits from standardized benchmarks; our framework benefits from production integration.

\paragraph{Limitation of this comparison.} We acknowledge that without running all frameworks on the same system under test, this comparison is structural rather than empirical. A rigorous comparative study---running multiple evaluation frameworks on the same agent and measuring their agreement with human judgments---would strengthen the evidence for the relative advantages of our approach. We leave this as future work.

\section{Conclusion}
\label{sec:conclusion}

We presented a three-layer dogfooding framework for evaluating LLM chat agents that integrates canonical testing, multi-turn depth evaluation, and goal-directed NPC simulation. Applied to a production system over roughly three months of active development, the framework demonstrates that:

\begin{enumerate}[leftmargin=*]
    \item The three layers provide \textbf{complementary regression signals}. No single layer catches all failure modes; canonical testing excels at structural failures while NPC simulation excels at conversational failures.
    \item Goal-directed NPC simulation with a \textbf{ten-category failure taxonomy} enables systematic diagnosis of conversational quality issues, directly informing targeted improvements.
    \item The framework is \textbf{cost-effective}: at \npcTotalCost{} per full evaluation cycle, it is \costRatio{} cheaper than human evaluation, enabling daily CI/CD integration.
    \item Dogfooding drives a \textbf{measurable improvement cycle}: evaluation $\rightarrow$ failure analysis $\rightarrow$ targeted fix $\rightarrow$ re-evaluation. Over our case study, the NPC pass rate improved from $\sim$57\% to 70\% through 10+ documented improvement phases, though the trajectory was non-monotonic (1 PROMOTE, 13 HOLD, 7 ROLLBACK across 21 runs).
\end{enumerate}

\paragraph{Future work.}
Three directions are particularly promising:
\begin{enumerate}[leftmargin=*]
    \item \textbf{Sim-to-real transfer validation:} Quantitatively correlating NPC failure patterns with real user satisfaction metrics (CSAT, NPS, support tickets) to establish predictive validity.
    \item \textbf{Multi-system replication:} Applying the framework to LLM chat agents in other domains (customer support, healthcare, education) to test generalizability.
    \item \textbf{Automated scenario generation:} Using production conversation traces to generate NPC scenarios that reflect real user behavior distributions, reducing authoring bias.
\end{enumerate}

The replicability guide, prompt templates, and failure taxonomy are provided to enable adoption by other teams building LLM chat agents.

\section*{Acknowledgments}
The author thanks two independent evaluators for their participation in the human calibration study. Neither evaluator is a co-author; their role was limited to blind response annotation according to the protocol described in Section~\ref{sec:results}.

\section*{AI Tools Disclosure}

This research leveraged AI-assisted development tools to support manuscript preparation and code development, while maintaining full human oversight and accountability. The following tools were used:
\begin{itemize}
    \item \textbf{Language models:} GPT-5 family (OpenAI via Codex),
    Claude Opus 4.7 (Anthropic Claude Code), and
    Google Gemini models (Gemini 2.5 Flash for system evaluation workloads, and
    Gemini 2.5 Pro as the independent LLM-as-judge in the human calibration study)
    were used to generate and review code implementations, and to refine manuscript text.

    \item \textbf{Web search:} MCP Tavily integration was used to support literature review and fact-checking during manuscript preparation.
\end{itemize}

All scientific arguments, empirical methodology, statistical analysis, research questions, and conclusions were independently conceived, developed, and validated by the author.

\section*{Statements and Declarations}
\paragraph{Funding}
This study did not receive a direct research grant. Experimental operation used Google Cloud resources, Gemini 2.5 Flash for system evaluation workloads, and Gemini 2.5 Pro for the LLM-as-judge calibration study.

\paragraph{Competing Interests}
The author declares no competing interests.

\paragraph{Author Contributions}
Single-author contribution. Alexandre Cristov\~ao Maiorano conceived the study, designed the methodology, implemented the analysis pipeline, conducted statistical analyses, interpreted results, and wrote and revised the manuscript.

\paragraph{Data Availability}
All aggregate results and statistical tables are reported in full in the paper
(Section~\ref{sec:results} and the appendices) and are auto-generated from the
underlying evaluation artifacts. The Layer~1 canonical-evaluation
artifacts---evaluation traces, question banks, the human-calibration protocol,
and judge-agreement data---are available in the companion replication
repository: \url{https://github.com/alemaiorano/dogfooding-bench}.
The Layer~3 NPC conversation traces, Layer~2 multi-turn traces, and production
system traces are not publicly released, as they contain proprietary
information about the production system under test; the paper reports all
aggregate metrics derived from them, and Appendix~\ref{sec:appendix-architecture}
provides the complete NPC scenario schema, prompt templates, and failure
taxonomy sufficient for independent reimplementation. Additional non-proprietary
methodological details may be provided by the author upon reasonable request.

\paragraph{Code Availability}
The companion replication repository provides the Layer~1 evaluation pipeline,
the LLM-judge and human-calibration scripts, and replication instructions.
The NPC simulator and multi-turn evaluation runners are part of the production
orchestration codebase and are not publicly released; Appendix~\ref{sec:appendix-replicability}
provides implementation-oriented pseudocode, data-contract schemas, and prompt
templates sufficient for independent adaptation.

\paragraph{Ethics Approval and Consent to Participate}
Not applicable.

\paragraph{Consent for Publication}
Not applicable.

\bibliographystyle{plainnat}
\bibliography{references}

\appendix
\section{NPC Simulator Architecture Details}
\label{sec:appendix-architecture}

\subsection{Scenario JSON Schema}

Each NPC scenario is defined by the following JSON structure:

\begin{lstlisting}
{
  "id": "scenario-unique-id",
  "persona": {
    "name": "Dr. Ricardo",
    "role": "team_lead",
    "knowledgeLevel": "expert",
    "communicationStyle": "technical",
    "language": "pt-BR"
  },
  "goal": {
    "type": "research_then_action",
    "description": "Find which option has the best performance, then apply it to the current project",
    "successCriteria": "Agent provides specific metric and executes the application"
  },
  "strategy": {
    "style": "thorough",
    "patience": 6,
    "driftTolerance": 0.4
  },
  "seedQuestion": "What are the metrics for each of my options?"
}
\end{lstlisting}

\subsection{NPC Generator Prompt Template}
\label{sec:appendix-prompts}

The NPC follow-up generator uses the following system prompt (model: Gemini 2.5 Flash, temperature 0.7). Template variables are shown in curly braces:

\begin{lstlisting}
You are role-playing as a real user interacting with an AI assistant.
You must stay in character throughout the conversation.

YOUR PROFILE:
- Persona: {persona}
- Communication style: {style}
- Knowledge level: {knowledgeLevel}
- Language: {language}

YOUR GOAL: {goalDescription}
Success condition: {successCondition}

CONVERSATION HISTORY:
{history}

STRATEGY:
- Style: {npcStyle} -- act accordingly:
  skeptical: challenge claims, demand proof
  curious: ask follow-ups, explore tangents
  impatient: short messages, want quick answers
  thorough: ask for details, examples, edge cases
  friendly: collaborative, positive, open
- Patience: {patience} (0-1).
- Remaining turns: {remainingTurns}.
- Stay focused on your goal but allow natural topic drift up to {driftTolerance}.

Generate the NEXT message. Consider:
1. Did the last response advance your goal?
2. What do you still need?
3. What's the most natural next step?
4. Would a real person with your profile continue or give up?

Return JSON:
{
  "text": "<your next message>",
  "intent": "<clarification|deeper_dive|comparison|action|challenge|pivot|give_up>",
  "goalProgress": "<none|partial|significant|complete>",
  "reasoning": "<brief explanation>"
}
\end{lstlisting}

\paragraph{Goal progress scale.}
\texttt{none}: no help toward the goal; \texttt{partial}: some progress but major gaps; \texttt{significant}: most addressed but details missing; \texttt{complete}: success condition met.

\subsection{NPC Judge Prompt Template}

The NPC judge evaluates completed conversations (model: Gemini 2.5 Pro, temperature 0.1):

\begin{lstlisting}
You are an expert evaluator assessing a multi-turn conversation
between a simulated user (NPC) and an AI assistant.

The NPC had a specific goal and persona. Evaluate whether the
conversation was successful.

Dimensions (0-5 scale):
- goalAchievement: Did the NPC achieve its goal?
  Score 4-5 if goalProgress was "complete".
  Score 0-1 if most turns show "none".
- personaConsistency: Did the NPC stay in character?
  Did the assistant adapt tone?
- conversationNaturalness: Did it feel real?
  Penalize robotic, repetitive phrasing.
- assistantHelpfulness: Was the assistant helpful and actionable?
  Penalize deflection.
- contextRetention: Did the assistant reference earlier turns?
  Score 0 if context ignored.

Pass: overall >= 3.0 AND goalAchievement >= 2.0.

Classify failures into:
context_loss, hallucination, routing_error,
goal_misalignment, persona_break,
generic_response, code_leakage, action_failure,
research_failure.

Return JSON:
{
  "goalAchievement": <0-5>,
  "personaConsistency": <0-5>,
  "conversationNaturalness": <0-5>,
  "assistantHelpfulness": <0-5>,
  "contextRetention": <0-5>,
  "overall": <0-5>,
  "pass": <true|false>,
  "failureCategories": ["<cat1>"],
  "reasoning": "<explanation>"
}
\end{lstlisting}

\paragraph{Note on taxonomy coverage in the judge prompt.}
The judge prompt enumerates nine of the ten taxonomy categories; \texttt{latency\_spike} is intentionally omitted because it is detected deterministically from response timers (Appendix~\ref{sec:appendix-taxonomy}, Table~\ref{tab:taxonomy-full}) rather than by the LLM judge.

\subsection{Data Contracts}

The framework produces three core JSON artifacts:

\paragraph{NPC Conversation Result.}
Each scenario produces a conversation result with the following structure:

\begin{lstlisting}
{
  "scenarioId": "unique-id",
  "goalType": "research|action|research_then_action|comparison|troubleshoot",
  "totalTurns": 4,
  "goalAchieved": true,
  "finalGoalProgress": "complete|significant|partial|none",
  "npcGaveUp": false,
  "failureCategories": [],
  "latencyMs": 12340,
  "judgeResult": {
    "goalAchievement": 4,
    "personaConsistency": 5,
    "conversationNaturalness": 4,
    "assistantHelpfulness": 4,
    "contextRetention": 4,
    "overall": 4,
    "pass": true,
    "reasoning": "..."
  },
  "cost": {
    "totalCost": 0.0015,
    "generatorTokens": 2450,
    "judgeTokens": 1800
  }
}
\end{lstlisting}

\paragraph{Aggregate Metrics.}
After all scenarios complete, aggregate metrics are computed:

\begin{lstlisting}
{
  "totalScenarios": 108,
  "passRate": 70,
  "goalAchievementRate": 77,
  "npcGiveUpRate": 19,
  "avgOverall": 3.66,
  "avgContextRetention": 4.01,
  "avgPersonaConsistency": 4.15,
  "avgConversationNaturalness": 3.78,
  "failureTaxonomy": {
    "goal_misalignment": 69,
    "generic_response": 54,
    "latency_spike": 41,
    "context_loss": 24,
    "persona_break": 21,
    "routing_error": 6,
    "research_failure": 5,
    "action_failure": 2,
    "hallucination": 2
  },
  "totalCost": 0.17,
  "gateResult": {
    "decision": "PROMOTE|HOLD|ROLLBACK",
    "gates": [{"name": "...", "value": 77, "threshold": 70, "passed": true}]
  }
}
\end{lstlisting}

\section{Replicability Guide}
\label{sec:appendix-replicability}

This appendix provides a condensed Python-first guide for replicating the three-layer evaluation framework on any LLM chat agent.

\subsection{Prerequisites}

\begin{itemize}[leftmargin=*]
    \item An LLM chat API endpoint (your system under test).
    \item An LLM API key for the judge (preferably a different model than the system).
    \item Python 3.10+ with \texttt{pandas}, \texttt{numpy}, and an LLM client library (\texttt{openai}, \texttt{google-genai}, or equivalent; our implementation uses the Gemini API).
\end{itemize}

\subsection{Step 1: Define NPC Scenarios}

Create JSON scenario files following the schema in Appendix~\ref{sec:appendix-architecture}. Start with 6--10 scenarios covering your most common user personas and goal types. Expand based on failure analysis.

\subsection{Step 2: Implement the Conversation Loop}

Algorithm~\ref{alg:npc-loop} specifies the core NPC--agent conversation loop, which alternates NPC follow-up generation with agent responses until the goal is achieved or the turn budget is exhausted.

\begin{algorithm}[H]
\caption{NPC Conversation Loop}\label{alg:npc-loop}
\begin{algorithmic}[1]
\State \textbf{Input:} scenario $s$, max turns $T$, agent API $A$
\State $\mathit{conv} \gets []$; $\mathit{turn} \gets 0$; $\mathit{gprog} \gets \text{none}$
\For{$\mathit{turn} < T$}
    \If{$\mathit{turn} = 0$}
        \State $\mathit{msg} \gets s.\mathit{seedQuestion}$
    \Else
        \State $\mathit{msg} \gets \textsc{GenFollowUp}(s, \mathit{conv}, \mathit{gprog})$
    \EndIf
    \State $\mathit{resp} \gets A.\textsc{Chat}(\mathit{msg}, \mathit{conv})$
    \State $\textsc{Append}(\mathit{conv}, (\mathit{msg}, \mathit{resp}))$
    \State $\mathit{gprog} \gets \textsc{AssessGoal}(s, \mathit{conv})$
    \If{$\mathit{gprog} = \text{complete}$}
        \State \textbf{break}
    \EndIf
\EndFor
\State \Return $\mathit{conv}, \mathit{gprog}$
\end{algorithmic}
\end{algorithm}

\subsection{Step 3: Score and Classify}

For each completed conversation, run the NPC Judge (6-dimension scoring) and classify failures using the ten-category taxonomy. Aggregate metrics across all scenarios and apply the quality gates.

\subsection{Step 4: Compute Bootstrap CIs}

\begin{lstlisting}[language=Python]
def bootstrap_ci(data, n_resamples=1000):
    stats = []
    for _ in range(n_resamples):
        sample = np.random.choice(
            data, size=len(data), replace=True)
        stats.append(np.mean(sample))
    return {
        "mean": np.mean(stats),
        "ci_95": np.percentile(stats, [2.5, 97.5]),
        "se": np.std(stats)
    }
\end{lstlisting}

\subsection{Step 5: Apply Quality Gates}

\begin{itemize}[leftmargin=*]
    \item Goal achievement rate $\geq 70\%$
    \item Judge overall $\geq 3.0$
    \item Context retention $\geq 3.0$
    \item Give-up rate $\leq 30\%$
\end{itemize}

Decision: all pass $\rightarrow$ PROMOTE; pass rate $\geq 60\%$ $\rightarrow$ HOLD; otherwise $\rightarrow$ ROLLBACK.

\section{Failure Taxonomy and Improvement Phases}
\label{sec:appendix-taxonomy}

\begin{table}[H]
\centering
\scriptsize
\begin{tabularx}{\columnwidth}{lXl}
\toprule
\textbf{Category} & \textbf{Detection Method} & \textbf{System Component} \\
\midrule
\texttt{context\_loss} & Judge score + heuristic & State management \\
\texttt{hallucination} & KB overlap check & Response agent \\
\texttt{routing\_error} & Agent path trace & Intent classifier \\
\texttt{latency\_spike} & Timer ($>$25s) & Infrastructure \\
\texttt{goal\_misalignment} & Judge score & Response agent \\
\texttt{persona\_break} & Judge score & NPC generator \\
\texttt{generic\_response} & 20+ regex patterns & FAQ agent \\
\texttt{code\_leakage} & JSON/code regex & Response agent \\
\texttt{action\_failure} & Judge + action trace & Action agent \\
\texttt{research\_failure} & Judge + source check & Research agent \\
\bottomrule
\end{tabularx}
\caption{Complete failure taxonomy with detection methods and responsible system components.}
\label{tab:taxonomy-full}
\end{table}

\begin{table}[H]
\centering
\scriptsize
\begin{tabularx}{\columnwidth}{llX}
\toprule
\textbf{Phase} & \textbf{Version} & \textbf{Root Cause \& Fix} \\
\midrule
2.5 & v5.5.0 & Repetition detection via embeddings; PT-BR KB creation; research fallback routing \\
2.6 & v5.6.0 & Scenario expansion 45$\to$69; factual grounding middleware; patience calibration \\
2.7 & v5.6.10 & Hardcoded FAQ prefix causing 40\% generic rate; sticky language; repetition retry \\
2.8 & v5.7.0-EXP25 & Reverted overly aggressive filter; brand leakage filter \\
2.9 & v5.7.0-EXP26 & No-data response builder for when KB lacks answer \\
2.10 & v5.7.0-EXP28 & Data query fallback; KB sections; 108-scenario expansion \\
\bottomrule
\end{tabularx}
\caption{Summary of NPC simulator improvement phases with root causes and fixes.}
\label{tab:phases}
\end{table}

\end{document}